\documentclass[aps,prc,showpacs,amssymb,amsmath,amsfonts,nofootinbib,floatfix]{revtex4-2}
\usepackage{graphicx}

\makeatletter
\def\p@subsection{}

\def\p@subsubsection{}
\makeatother
\usepackage{natbib}
\usepackage[usenames]{color}
\usepackage{epsfig,graphicx}
\usepackage{epstopdf}
\usepackage{amssymb,amsmath,eqparbox}
\usepackage{subfig}
\usepackage{epstopdf}
\usepackage{enumitem}
\usepackage[bookmarks={false}]{hyperref}
\usepackage{enumitem}
\usepackage{bbm}

\definecolor{grey}{rgb}{0.9,0.9,0.9}
\definecolor{black}{rgb}{0,0,0}

\hypersetup{pdfstartview={XYZ null null 1.}}

\def\thesection{\Roman{section}}
\def\thesubsection{\Roman{section}. \Alph{subsection}}
\def\thesubsubsection{ \textit{\textbf{\Roman{section}. \Alph{subsection}. \arabic{subsubsection}}}}

\newcommand{\be}{\begin{eqnarray}}
\newcommand{\ee}{\end{eqnarray}}
\newcommand{\bc}{\begin{center}}
\newcommand{\ec}{\end{center}}

\begin{document}

\title{Is the pole content of each single-energy, single-channel partial wave analysis \\ inherently model dependent? }
\email{Corresponding author: alfred.svarc@irb.hr}
\author{ A. \v{S}varc\,$^{1,2}$ \vspace*{0.3cm}  }

\affiliation{$\,^1$ Rudjer Bo\v{s}kovi\'{c} Institute, Bijeni\v{c}ka cesta 54, P.O. Box 180, 10002 Zagreb, Croatia}
\affiliation{$\,^2$ Tesla Biotech d.o.o., Mandlova 7, 10000 Zagreb, Croatia  \vspace*{0.3cm}}


\date{\today }

\begin{abstract}
\vspace*{0.5cm}
Unfortunately, yes it is. In ref.\cite{Svarc2018} it has been shown that without fixing the reaction-amplitude phase, many combinations of partial waves at neighbouring energies in single-energy, single-channel partial wave analysis reproduce experimental data identically, but are discontinuous and disconnected. To obtain the continuous solution,  the phase has to be fixed to some continuous value. In the same reference it has also been shown that the change of angular part of reaction-amplitude phase mixes partial waves, so the pole structure of any single-energy single-channel partial wave analysis depends on the chosen phase.  As in any single-channel analysis the overall reaction-amplitude phase  cannot be determined because of continuum ambiguity,  it is in principle free and has to be taken from some coupled-channel model. Because of the difference in the angular part of the phase, choosing different phases results in the change of the pole content of the obtained solution. Therefore, single-energy single-channel partial wave analysis is inherently model dependent, and the number of poles it contains strongly depends on the choice of the phase. In that reference these facts have been illustrated on the pseudo-scalar meson production toy model.  This truth is  now for the measured observables demonstrated on the  realistic model of ref.~\cite{Svarc2020} which combines amplitude and partial wave analysis in one self-consistent scheme for the the world collection of $\eta$-photoproduction data, and fixes the phase to BG2014-2 solution \cite{BoGa}. We show that the pole structure of the solution whose phase is very close to the BG2014-2 theoretical ED phase is very similar to the pole structure of BG2014-2 solution, but the agreement with the data is not ideal. However, improving the agreement with the data requires the departure from BG2014-2 ED phase, and this spoils analytic structure of the obtained solution. This also demonstrated under which conditions the new method of ref.~\cite{Svarc2020} is reliable for direct data analysis and generating single-channel, single-energy partial wave analysis with the minimal model dependence, and we offer a new solution which is a compromise between good fit and good analytic structure.
\end{abstract}

\maketitle
\vspace*{-0.5cm}
\section{Introduction}
In ref.~\cite{Svarc2020}  we have formulated a single channel, single-energy partial wave analysis (SC-SE-PWA)  model which combines amplitude and partial wave analyses into one logical sequence, and directly from the data generates a set of continuous partial waves with minimally model dependent input~(AA/PWA). We have demonstrated that by controlling  the reaction-amplitude phase, and freely varying the reaction-amplitude partial waves,  we obtain a continuous solution with far better agreement with the used data base than the original ED model. We have generated two solutions,  (Sol 1) when the phase is close, but not identical to the phase of a particular coupled-channel, energy-dependent model (ref.~\cite{BoGa} in our case), and the second one in which the phase has been freely smoothed to a similar but simpler analytic function (Sol 2). We have shown that the quality of both fits is fairly similar, but the obtained higher multipoles are rather different ($E0^+$ is always dominantly big and very stable in $\eta$-photoproduction).  We have also promised to analyze their analytic structure (pole content) using Lauren+Pietarinen (L+P) technique~\cite{Svarc2016}. In this paper we fulfill the given promise for Sol 1. However, in doing so, some unexpected effects revealed themselves. The pole content of Sol 1 turned out to be unclear. In looking for the explanation why this is so, we discovered the illuminating truth of how and how much the precision in the phase determination decides on the analytic structure of the obtained solution. The main aim of this article is to show that the analytic structure of the AA/PWA model of ref.~\cite{Svarc2020} is correct, that the change of analytic structure (pole content)  of the obtained solution  depends on the changes of the phase, and quantify when the notable deviation in the pole content starts. This will decide how well we can fit the given data base at the same time maintaining the proper analytic structure (pole content) determined by the constraining ED model. Further improvements in the fit without controlled phase change will be able only by spoiling good pole content, and introduce ghost poles originating in the change of the angular part of the phase. Therefore we show that the only possible improvement in present SC-SE-PWA is also model dependent, and it is accomplished  by improving the constraining phase of the coupled-channel ED model. Free change of phase is not allowed, and it directly leads to  ghost poles.  As a direct corollary of this conclusion the analysis of Sol 2 turned out to be unnecessary\footnote{Sol 2 was obtained with free smoothing of the reaction amplitude phase not controlled by other channels, so appearance of ghost poles which spoil the analytic structure is unavoidable.}.
\\ \\ \indent
 To understand that, we have discuss the problem of analytic structure of obtained solutions in the wider context of angular dependent phase rotations.
\\ \\ \indent
 In ref.~\cite{Svarc2018} it has been shown that without fixing the reaction-amplitude phase, many combinations of partial waves at neighbouring energies in single-energy single-channel partial wave analysis reproduce experimental data identically, but are discontinuous and disconnected. So, to obtain the continuous solution,  the phase has to be fixed to some continuous value. In the same reference it has also been shown that the change of angular part of reaction-amplitude phase mixes partial waves so the pole structure of any single-energy single-channel partial wave analysis depends on the chosen phase.  As the reaction-amplitude phase because of continuum ambiguity cannot be determined in any single-channel analysis, it has to be taken from some coupled-channel model. Choosing different phases results in the change of the pole content of the obtained solution. Therefore, SC-SE-PWA must depend on the reaction-amplitude phase, and it has to be taken from some model. Hence SC-SE-PWA must be inherently model dependent. All these statements were demonstrated on the toy model of pseudo-scalar meson production.
  \\ \\ \indent
  In this paper this fact is quantitatively demonstrated on the real data in the single-channel single-energy model AA/PWA  of ref.~\cite{Svarc2020} which combines amplitude and partial wave analysis in one self-consistent, 2-step model for the the world collection of data in $\eta$-photoproduction. We first perform the amplitude analysis (AA) of the world collection of data fixing the amplitude phase to the theoretical ED BG2014-2 phase and varying only absolute values as the free parameters. In the second step we perform a constrained truncated partial wave analysis (TPWA) where multipoles for the L=0 to L=5 are free parameters, but the fit was constrained by requiring that the final reaction amplitudes do not differ much from the AA amplitudes of Step 1. In this way we achieve a continuity of otherwise discontinuous multipoles. Let us immediately stress that the phase of the final solution depends strongly  on the amount of penalization. The higher the penalization is, the closer the phase of the final solution is to the chosen theoretical ED phase.  However, we have discovered that the quality of the agreement of the final fit with the data ($\chi^2$)  notably depends on the amount of penalization. If the penalization is only moderate we obtain notably better agreement with the data than in the case when penalization is strong. However, the price paid is that for the better  $\chi^2$ with the milder penalization the departure of the final phase from the theoretical BG2014-2 phase is bigger. Next step was finding poles of the obtained solution. As we are dealing with SC-SE-PWA, the only option is using L+P formalism. As we start from BG2014-2 ED solution, we have first extracted poles from ED multipoles using L+P formalism. We have shown that the extracted poles, qualitatively and quantitatively, perfectly correspond to published values obtained by analytic continuation of theoretical coupled-channel amplitudes into the complex energy plane. Then, using the same L+P model (same number of Pietarinen coefficients, same number of variable parameters...) we have searched for the poles of our Sol 1, whose phase is much closer to the ED phase than for Sol 2. Poles turned out to be much less stable, and notably different from the poles extracted from ED solution. We attributed this instability to spoiling the phase with respect to the input ED phase. To test this hypothesis we have repeated the AA/PWA model with much stronger penalty (we have increased the penalty function coefficient from $\lambda_{penalty}$=10 for ref.~~\cite{Svarc2020} to $\lambda_{penalty}$=500  for this publication), and obtained a new soution Sol 1/21 which is a compromise between a good fit and correct analytic structure. As expected, the overall agreement with the data was slightly spoiled, but the analytic structure of the new solution was notably improved. And this was what we wanted to prove.
  \section{Formalism}
  In this paper we use the improved formalism of AA/PWA formulated in ref.~\cite{Svarc2020}. As PWA is without constraining conditions inherently discontinuous, leaded by the idea of fixed-t analyticity of ref.~\cite{Hoehler84}  we have formulated a two step procedure which in the second step constrains the PWA with reaction amplitudes obtained in the AA from the previous step. As the phase because of continuum ambiguity cannot be determined out of the data in any single-channel analysis, we had to take the phase from well know coupled-channel theoretical model, and the  BG2014-2 \cite{BoGa} model was chosen. In the AA of the original publication only four out of eight available observables were fitted with absolute values as free parameters, and these were phase independent observables $d\sigma/d\Omega$, $\Sigma$, $T$ and $P$. This actually was the exact single energy amplitude reconstruction (4 observables for 4 absolute values). In this publication we have extended the number of observables to all eight available observables, so this was not an exact amplitude reconstruction, but the actual fit. The phase is, as before, fixed to BG ED phase. In the ideal case of an  infinitely precise complete experiment, fixing the phase is enough to obtain unique amplitudes, continuous in both, energy and angle. However, as the existing data base is far from a complete and self-consistent data set, we enforced the continuity by penalizing the fit with the obtained result at neighbouring energies\footnote{It is possible as this is a single-energy analysis, meaning that the minimization is performed independently at each chosen energy}. In this way the set of continuous reaction amplitudes was obtained, and all deviations from the perfectly smooth analytic function were due only to experimental errors. The next step was to obtain corresponding multipoles. In the ideal case of a complete set of exact observables available at sufficiently dense energies, multipoles extraction is trivial, and boils down to integral over Legendre polynomials of a proper combination of amplitudes. However, due to incompleteness of the available data base this is not possible. So, we have applied penalty function technique where the PWA of the data is penalized with the result of AA step. And this is the step where the departure in phase of our final result from the initially used ED phase occurs. As it turns out, the agreement of the obtained result with the data in the AA step is not perfect, and better agreement with the data can be obtained in constrained PWA. Our analysis of the penalty function of Step 2 shows that the improvement in $\chi^2$ is achieved mainly because of deviation of the phase of the final solution from the input phase which is fixed in the AA step.

  For the convenience of the reader we repeat the essential equations governing the two step AA/PWA method from ref.~\cite{Svarc2020}.
  The most standard, classic approach was to penalize partial waves by requiring that fitted partial waves reproduce the observable $\cal O$   and are at the same time close to some partial waves taken from the theoretical model:
\be
\label{Eq1}
\chi^2(W) & = & \sum_{i=1}^{N_{data}}w^i \left[ {\cal O}^{exp}_i (W,\Theta_i) - {\cal O}^{th}_i ({\cal M}^{fit}(W,\Theta_i)) \right]^2 + \lambda_{pen} \sum_{i=1}^{N_{data}}\left[ {\cal M}^{fit}(W,\Theta_i)- {\cal M}^{th}(W,\Theta_i) \right]^2
\ee
where
\be
{\cal M} & \stackrel{def}{=} & \left\{ {\cal M}_0, {\cal M}_1, {\cal M}_2, ..., {\cal M}_j \right\} \nonumber
\ee
$w_i$   is the statistical weight and $j$ is the number of  partial waves (multipoles). Here  ${\cal M}^{fit}$  are fitting parameters and ${\cal M}^{th}$ are continuous functions taken from a particular theoretical model.
Instead, the possibility to make the penalization function independent of a particular model was first formulated in Karlsruhe-Helsinki $\pi$N elastic PWA by G. H\"{o}hler in mid-80es~\cite{Hoehler84}.  Partial waves which are inherently model dependent are replaced with the penalization function which was constructed from reaction amplitudes which can be in principle directly linked to experimental data without any model in amplitude reconstruction procedure. So, the equation \ref{Eq1} was changed to:
\be
\label{Eq2}
\chi^2(W) & = & \sum_{i=1}^{N_{data}}w^i \left[ {\cal O}^{exp}_i (W,\Theta_i) - {\cal O}^{th}_i ({\cal M}^{fit}(W,\Theta_i)) \right]^2
+ \cal{P}  \\ \nonumber
 \cal{P}   &=& \lambda_{pen} \sum_{i=1}^{N_{data}} \sum_{k=1}^{N_{amp}}\left| {\cal A}_k({\cal M}^{fit}(W,\Theta_i))- {\cal A}_k({\cal M}^{pen}(W,\Theta_i)) \right|^2
\ee
where ${\cal A}_k$  is the generic name for any of reaction amplitudes (invariant, helicity, transversity…). However, one is now facing two challenges: to get reaction amplitudes which fit the data, and also to make them continuous. In Karlsruhe-Helsinki case it was accomplished by implementing fixed-t analyticity and fitting the data base for fixed-t.  So, the first step of the KH fixed-t approach was to create the data base ${\cal O}(W)|_{t=fixed}$  using the measured base $ {\cal O}(\cos \, \theta)|_{W=fixed}$, and then to fit them with manifestly analytic representation of reaction amplitudes for a fixed-t. Manifest analyticity was implemented by using Pietarinen  decomposition of reaction amplitudes.  Then the second step was to perform a penalized PWA defined by Eq.~\ref{Eq2} in fixed-W channel where the penalizing factor  ${\cal A}_k({\cal M}^{pen}(W,\Theta_i)) $  was obtained in the first step in a fixed-t channel.  In that way a stabilized SE PWA was performed. This approach  was revived very recently for SE PWA of $\eta$ and $\pi^0$-photoproduction by Main-Tuzla-Zagreb collaboration, and analyzed in details in refs.~\cite{Osmanovic2018,Osmanovic2019}.
\\ \\
We propose the alternative.
\\ \\
We also use Eq.~\ref{Eq2}, but the penalizing factor ${\cal A}_k({\cal M}^{pen}(W,\Theta_i)) $ is  generated by the amplitude analysis in the same, fixed-W representation, and not in the fixed-t one. This simplifies the procedure significantly, and avoids quite some theoretical assumptions on the behaviour in the fixed-t representation.
\\ \\
We also propose a 2-step process as in ref.~\cite{Hoehler84,Osmanovic2019}:
\begin{itemize}[leftmargin=2.cm]
  \item[\emph{Step 1:}] \hspace*{1.cm} \\
   Amplitude analysis of experimental data in fixed-W system to generate penalizing factor  ${\cal A}_k({\cal M}^{pen}(W,\Theta_i)) $
  \item[\emph{Step 2:}] \hspace*{1.cm} \\
   Penalized PWA Using Eq.(2) with the penalization factor from
         \emph{Step 1}.
\end{itemize}

    However, in this paper we improve the procedure of ref.~\cite{Svarc2020}. In that reference  for the \emph{Step 1} (AA) only the four phase independent observables $d\sigma/d\Omega$, $\Sigma$, $T$ and $P$ were fitted obtaining for absolute values as exact amplitude reconstruction\footnote{We use transversity amplitude representation.}, and here we extend the process to all eight available observables. The phase is in both cases fixed to theoretical BG ED phase. In this way we  get the best set of transversity amplitudes which maximally reproduce all available observables for the given phase. However, this is where the model dependence starts. Transversity amplitude phase of constraining amplitudes is fixed, so agreement of the fit for the four phase dependent observables $E$, $F$, $G$, and $H$ can be improved only through the absolute value but they are also phase dependent, so it is limited. Possibility of further improvement for these observables can only be achieved in \emph{Step 2}, constrained PWA. It is clear that the amount of departure from the BG ED phase depends on the level of penalization. Smaller penalization means bigger departure from the BG phase, the $\chi^2$ for the phase dependent data improves, but partial waves get less continuous and  mixed. As it will be shown later in this paper, this is clearly felt by the L+P pole extraction method. For the strong penalization the analytic structure of multipoles (pole content) strongly resembles the analytic structure corresponding to BG ED model (and PDG therewith), for the weaker penalization the position of poles gets much less precise.
\\ \\   \indent
  And now we are bound to say something about the importance of the reaction amplitude phase. Continuum ambiguity forbids to conclude onto the correct phase in any single channel analysis because unitarity loss to other channels starts after the first inelastic threshold opens. The only way to solve the continuum ambiguity problem is to reintroduce the unitarity introducing coupled-channel formalism. If we pick the phase in a single-channel analysis arbitrarily by hand, we are departing  from the genuine phase, the phase in which partial waves do not mix, and we via angular part of continuum ambiguity introduce pole transfer from one partial wave into another. However, each coupled-channel model by construction results in the non-mixing pole solution. Namely, some form of interaction introducing poles is formulated,  and the background contribution is added to it. Then, the data in all channels are simultaneously fitted forcing the phase to be the correct one, and the pole non-mixing situation is established. Background contributions automatically enforce the phase to be a non-mixing one. It is needless to say that all coupled-channel models should end up with the same phase in the ideal case, but incompleteness of the data forbids that to happen. Therefore, phases of different models \cite{BoGa,Juelich,MAID,GWU/SAID} are somewhat different, and we cannot avoid this. However, fixing the phase to a phase of a particular model ensures to obtain the non-mixing pole solution; departure from it automatically enforce pole mixing, so the analytic structure of such a solution is spoiled. So, we can chose a different phase, a phase coming from any model, but it has to be the proper phase originating from that model. Free, uncontrolled departure from ED model phase is not allowed.
\\ \\   \indent
  Of course the question is purely quantitative: How much can we depart from the "diagonal" phase in an uncontrolled way to maintain the correct analytic property. In other words, the question is how much we are allowed to reduce the importance of the penalty function and main the correct analyticity.
\newpage
\section{Results and Discussion}

\subsection{Analytic Structure (Pole Content) of Sol 1}
As we are interested in analyzing the pole structure of the obtained solution Sol 1 which is the SE quantity, we use the most sophisticated method which enables pole extraction from SE quantities, and that is the Laurent+Pietarinen (L+P) formalism \cite{Svarc2016}. This, relatively recent model is based on fitting, instead of exact mathematical analytic continuation methods. The assumption is simple:  we do not construct an analytic function on the real axes, and  continue it into the complex energy plane, but instead using the most general principles we construct one in the complex energy plane in the vicinity of the real axes, and fit the free parameters to the data on the real axes. The method entirely relies on analyticity. As we assume that each physical process must be described by an analytic function, we start with its Laurent decomposition. We know that by Laurent theorem any analytic function is locally,  within limited area of convergency, uniquely determined with its poles and cuts, and cuts are generated by channel opening branch-points.  For the singular part we  therefore have 4 parameters per pole, and for the regular part we use the fast converging expansion over conformal variables generated by a chosen number of relevant branch-points. As the analytic function is entirely determined by its poles and cuts, by choosing enough poles and cover all relevant branch-points, our solution represents the simplest analytic function which corresponds to the data on the real axes. The simplicity is enforced by choosing the lowest number of  free parameters for poles and cuts which are then fitted to the data.


The model is standard, coupled-multipole model of ref.~\cite{Svarc2016}. We use three Pietarinen expansions, first and third branch-points are left free. The middle branch-point is fixed to $\eta$-photoproduction threshold.  The number of Pietarinen terms is limited to 3-5 per expansion, and starting number of poles is set to the accepted 3,4-star PDG resonances for the given multipole.

First we use the described L+P model to extract pole parameters from BG 2014-2 solution of ref.~\cite{BoGa} to obtain the reference point, and then from Sol 1 of ref.~\cite{Svarc2020}.
Extracting poles from BG 2014-2 multipoles is straightforward, results are very confident, and completely correspond to the values given in the literature \cite{BoGa,PDG}. We can compare only pole positions with the literature, residues for $\eta$-photoproduction process are for the first time extracted in this publication.   The results are give in Table~\ref{Table1}. It is clear that pole parameters extracted from Sol 1 very poorly  match the values extracted from BG 2014-2 solution, and this should not be so. Therefore, we have to find the explanation.
\newpage

\subsection{Analysis of Phase Dependence of Sol 1}
In Table~\ref{Table1} we present all results obtained for poles using L+P analysis.

\begin{table}[htb]
\begin{center}

\caption{\label{Table1} Pole parameters for BG 2014-2, Sol 1 and Sol 1/21 extracted using L+P expansion }
\bigskip
\scalebox{0.95}{
\begin{tabular}{cccccccccccc} \hline \hline  \\[-2ex]
    & Model    & $M_1$ & $\Gamma _1$ & $|a_1|$  & $\Theta_1$ & $M_2$ & $\Gamma _2$ & $|a_2|$  & $\Theta_2$ & $\chi^2$ & $\chi^2_{red}$    \\[0.5ex]
\hline  \\[-2ex]
  $S_{11}$  $1/2^-$    &  PDG    &  \textbf{1510(19)}  &     \textbf{ 130(20)}      &     -     &      -       &  \textbf{1655(15)}     &    \textbf{135(35)}          &    -      &   -          &          &  \\
    &  BG 2014-2    &  1498(107)     &  158(157)    &  1780(5300)     &   164(345)   &  1661(5)    &  85(12)    &   126(47)       &   24(19)          &   18       &  0.13  \\
  $E_0^+$    &  Sol 1    &    1489(66)    &    158(78)        &     2043(5054)     &     148(146)       &    1664(5)    &            92(9) &      140(36)    &    37(15)         &  140        &  0.6 \\
              &  Sol 1/21    &   1484(37)    &     196(189)       &  2926(7330)    &     166(172)    &  1662(3)    & 101(7)    &    158(26)      &   34(9)          &    97      &  0.43 \\
\hline \\[-2ex]
   $P_{11}$  $ 1/2^+$    &  PDG    &  \textbf{1379(10) }  &    \textbf{175(15) }    &     -     &      -       &  \textbf{1700(20) }   &   \textbf{  120(40)}         &    -      &   -          &          &  \\
      &  BG 2014-2    &  -  &  -   &  -   &  -  &  1698(1)      &  123(1)   &    105(2)       &    -90(1)          &   102       &  0.75  \\
   $M_1^-$   &  Sol 1    &  -    &  -     &  -       &   - &    1730(6)    &   80(10)     &   48(12)       &     -22(18)          &  140        &  0.7   \\
              &  Sol 1/21    &   -    &     -      &  -  &     -   & 1660(6)    &     112(13)       &  49(15))    &     -168(16)     &    106   &  0.47  \\
              &  Sol 1/21    &   1526(25)   &     73(37)      &  19(32)  &     -123(110)   & 1681(7)    &     103(12)       &  39(10))    &     -124(17)     &    25   &  0.11 \\
\hline \\[-2ex]
   $P_{13}$  $ 3/2^+$    &  PDG    &  - &   -    &     -     &      -       &  \textbf{1675(15) }   &   \textbf{  250(150)}         &    -      &   -          &          &  \\
      &  BG 2014-2    &  -  &  -   &  -   &  -  &  1705(7)      &  195(21)   &      $\left( ^{\eqmakebox[aaa][c]{38(10)}}_{\eqmakebox[aaa][c]{38(13)}} \right)$       &  $\left( ^{\eqmakebox[aaa][c]{-133(15)}}_{\eqmakebox[aaa][c]{-107(16)}} \right)$    &      58       &  0.23  \\
 \large{$\left( ^{E_1^+}_{M_1^+} \right)$}     &  Sol 1    &  -    &  -     &  -       &   - &    1879(46)    &   200(68)     &  $\left( ^{\eqmakebox[aaa][c]{328(359)}}_{\eqmakebox[aaa][c]{260(280)}} \right)$  &   $\left( ^{\eqmakebox[aaa][c]{-8(44)}}_{\eqmakebox[aaa][c]{ 67(50)}} \right)$   &     310        &  0.6   \\
          &  Sol 1/21    &   -    &     -      &  -  &     -   & 1714(7)    &     102(13)       & $\left( ^{\eqmakebox[aaa][c]{10(3)}}_{\eqmakebox[aaa][c]{ 1(1)}} \right)$    &     $\left( ^{\eqmakebox[aaa][c]{-167(16)}}_{\eqmakebox[aaa][c]{ 20(47)}} \right)$    &    297   &  0.6  \\
 \hline \\[-2ex]
 $D_{13}$  $ 3/2^-$             &  PDG    &  \textbf{1510(5) } &   \textbf{110(10) }    &     -     &      -       &  \textbf{1700(50) }   &   \textbf{  200(100)}         &    -      &   -          &          &  \\
      &  BG 2014-2    &  1508(3)  &  106(7)   &   $\left( ^{\eqmakebox[aaa][c]{52(11)}}_{\eqmakebox[aaa][c]{25(6)}} \right)$    &  $\left( ^{\eqmakebox[aaa][c]{122(12)}}_{\eqmakebox[aaa][c]{118(13)}} \right)$  &  1664(76)      &  399(159)   &      $\left( ^{\eqmakebox[aaa][c]{119(155)}}_{\eqmakebox[aaa][c]{72(86)}} \right)$       &  $\left( ^{\eqmakebox[aaa][c]{73(71)}}_{\eqmakebox[aaa][c]{103(77)}} \right)$    &     1.7       &  0.06  \\
 \large{$\left( ^{E_2^-}_{M_2^-} \right)$}     &  Sol 1    &  1528(23)   &  63(37)     &  $\left( ^{\eqmakebox[aaa][c]{11(22)}}_{\eqmakebox[aaa][c]{2(3)}} \right)$        &   $\left( ^{\eqmakebox[aaa][c]{-160(82)}}_{\eqmakebox[aaa][c]{148(98)}} \right)$  &    1721(6)    &   64(13)     &  $\left( ^{\eqmakebox[aaa][c]{10(3)}}_{\eqmakebox[aaa][c]{4(1)}} \right)$  &   $\left( ^{\eqmakebox[aaa][c]{149(19)}}_{\eqmakebox[aaa][c]{ -168(18)}} \right)$   &    368        &  0.8   \\
          &  Sol 1/21    &   1525(23)    &     121(60)     &  $\left( ^{\eqmakebox[aaa][c]{37(52)}}_{\eqmakebox[aaa][c]{ 24(39)}} \right)$    &     $\left( ^{\eqmakebox[aaa][c]{-156(91)}}_{\eqmakebox[aaa][c]{ 158(94))}} \right)$     & 1664(12)    &     121(24)       & $\left( ^{\eqmakebox[aaa][c]{11(6)}}_{\eqmakebox[aaa][c]{ 13(7)}} \right)$    &     $\left( ^{\eqmakebox[aaa][c]{-31(33)}}_{\eqmakebox[aaa][c]{ 46(33)}} \right)$    &    50   &  0.1  \\
\hline \\[-2ex]
 $D_{15}$  $ 5/2^-$             &  PDG    &  - &   -    &     -     &      -       &  \textbf{1660(5) }   &   \textbf{  135(15)}         &    -      &   -          &          &  \\
      &  BG 2014-2    &  -  &  -   &  -   &  -  &  1673(4)      &  225(6)   &      $\left( ^{\eqmakebox[aaa][c]{1(0.3)}}_{\eqmakebox[aaa][c]{23(1)}} \right)$       &  $\left( ^{\eqmakebox[aaa][c]{54(17)}}_{\eqmakebox[aaa][c]{-17(6)}} \right)$    &      45       &  0.16  \\
 \large{$\left( ^{E_2^+}_{M_2^+} \right)$}     &  Sol 1    &  -    &  -     &  -       &   - &    1784(1)    &   11(1)     &  $\left( ^{\eqmakebox[aaa][c]{0.5(0.01)}}_{\eqmakebox[aaa][c]{0.8(0.1)}} \right)$  &   $\left( ^{\eqmakebox[aaa][c]{23(9)}}_{\eqmakebox[aaa][c]{ 95(9)}} \right)$   &     310        &  0.6   \\
          &  Sol 1/21    &   -    &     -      &  -  &     -   & 1659(10)    &     145(23)       & $\left( ^{\eqmakebox[aaa][c]{6(2)}}_{\eqmakebox[aaa][c]{ 9(4)}} \right)$    &     $\left( ^{\eqmakebox[aaa][c]{17(20)}}_{\eqmakebox[aaa][c]{ -40(22)}} \right)$    &    90   &  0.2  \\
 \hline \\[-2ex]
  $F_{15}$  $ 5/2^+$            &  PDG    &  - &   -    &     -     &      -       &  \textbf{1675(10) }   &   \textbf{  120(15)}         &    -      &   -          &          &  \\
      &  BG 2014-2    &  -  &  -   &  -   &  -  &  1677(1)      &  117(1)   &      $\left( ^{\eqmakebox[aaa][c]{13(1)}}_{\eqmakebox[aaa][c]{7(0.5)}} \right)$       &  $\left( ^{\eqmakebox[aaa][c]{147(1)}}_{\eqmakebox[aaa][c]{145(11)}} \right)$    &      12       &  0.05  \\
 \large{$\left( ^{E_3^-}_{M_3^-} \right)$}     &  Sol 1    &  -    &  -     &  -       &   - &    1767(2)    &   34(4)     &  $\left( ^{\eqmakebox[aaa][c]{3(0.5)}}_{\eqmakebox[aaa][c]{2(0.5)}} \right)$  &   $\left( ^{\eqmakebox[aaa][c]{-91(9)}}_{\eqmakebox[aaa][c]{ 33(10)}} \right)$   &     690        &  1.5   \\
          &  Sol 1/21    &   -    &     -      &  -  &     -   & 1690(4)    &     166(11)       & $\left( ^{\eqmakebox[aaa][c]{11(2)}}_{\eqmakebox[aaa][c]{ 23(4)}} \right)$    &     $\left( ^{\eqmakebox[aaa][c]{172(8)}}_{\eqmakebox[aaa][c]{ 164(77)}} \right)$    &    156   &  0.34  \\[1ex]
 \hline \hline
 \end{tabular} }
\end{center}
\end{table}
First we see that nice and stable poles are obtained when L+P is used on BG2014-2 model. However, we also immediately see some problems which will be blown up by additional ambiguities in SE analysis. First, the most stable part of the L+P formalism are pole positions. This is not at all unexpected as this is one of the main features of L+P formalism. The main model dependence of L+P formalism lies in pole-background separation. This part is rather arbitrary. The background is completely unknown, and is practically fitted to the input data, so a direct consequence is that residues of the singular part must strongly depend on the way how pole-background separation is performed (every change in background part is counterbalanced by the change in residue). On the other hand, pole position (position of the singularity in the complex energy plane) is for all possible backgrounds always the same. So, a direct consequence is that residues are much less precisely determined in L+P formalism, and may have some meaning only in the analysis of ED functions. There exists another reason for the uncertainty in residues in inelastic processes like $\eta$-photoproduction, and this is the ambiguity in reaction amplitude phase. As the phase can be determined only when multi-channel unitarity is restored, phase in singe-channel analyses is unknown. We also know that change in angular part of reaction amplitude phase mixes multipoles, so residues associated with a certain pole also depend on the chosen phase. So, this is another source of ambiguity for the residues.\footnote{Observe that this is not so in elastic processes like $\pi N$  elastic scattering as the unitarity gets violated only above first inelastic threshold. So the phase is up to this energy fully determined. First uncertainties occur at higher energies. However, let us observe that the phase of inelastic process like $\pi$-photoproduction is also fairly well determined at lower energies because of Watson's theorem which connects the $\pi$-photoproduction phase with $\pi N$ elastic phase at lower energies. Uncertainties also rise with energy, but are still small in the $N^*$ energy range.} Because of these two reasons, in the analysis of SE data, residues are  something just a little bit more than pure fitting parameters. This is obvious from Table~\ref{Table1}. Even for ED input, confidence level of residues is very low, sometimes on the level of 100 \% (for $S_{11}$ partial wave even much lower for the first resonance, but this is altogether another story\footnote{First $S_{11}$ pole is very near the threshold of $\eta$-photoproduction which is taken as the fixed branch-point in L+P formalism, so to have some impact onto the process the residue in addition to being unprecise must also be very high. However, this is typical even for ED models \cite{MAID}.}). However, poles are pretty precisely determined. Even for pole position determination, one can see a clear hierarchy in confidence level: pole position is always more precisely determined that its width (how deep the pole is in the complex energy plane is more unreliable). This is the direct consequence of the fact that all the fits are done with  the data lying on the real axes of the complex energy plane, so it is more sensitive to the position which is more directly influencing shape of the scattering matrix on the real axes.

Making an L+P analysis of the SE Sol 1, and even taking these facts into account, we see that the analytic structure of Sol 1 is unclear, and clearly rather different from the the ED solution we have started with. With the exception of $S_{11}$ all poles are notably shifted in energy what is not allowed, and pole width is unreasonably narrow. This is not admissible, so there must be some effect which spoils the analytic structure of Sol 1.

\begin{figure}[h!]
\bc
\includegraphics[height=0.25\textwidth,width=0.4\textwidth]{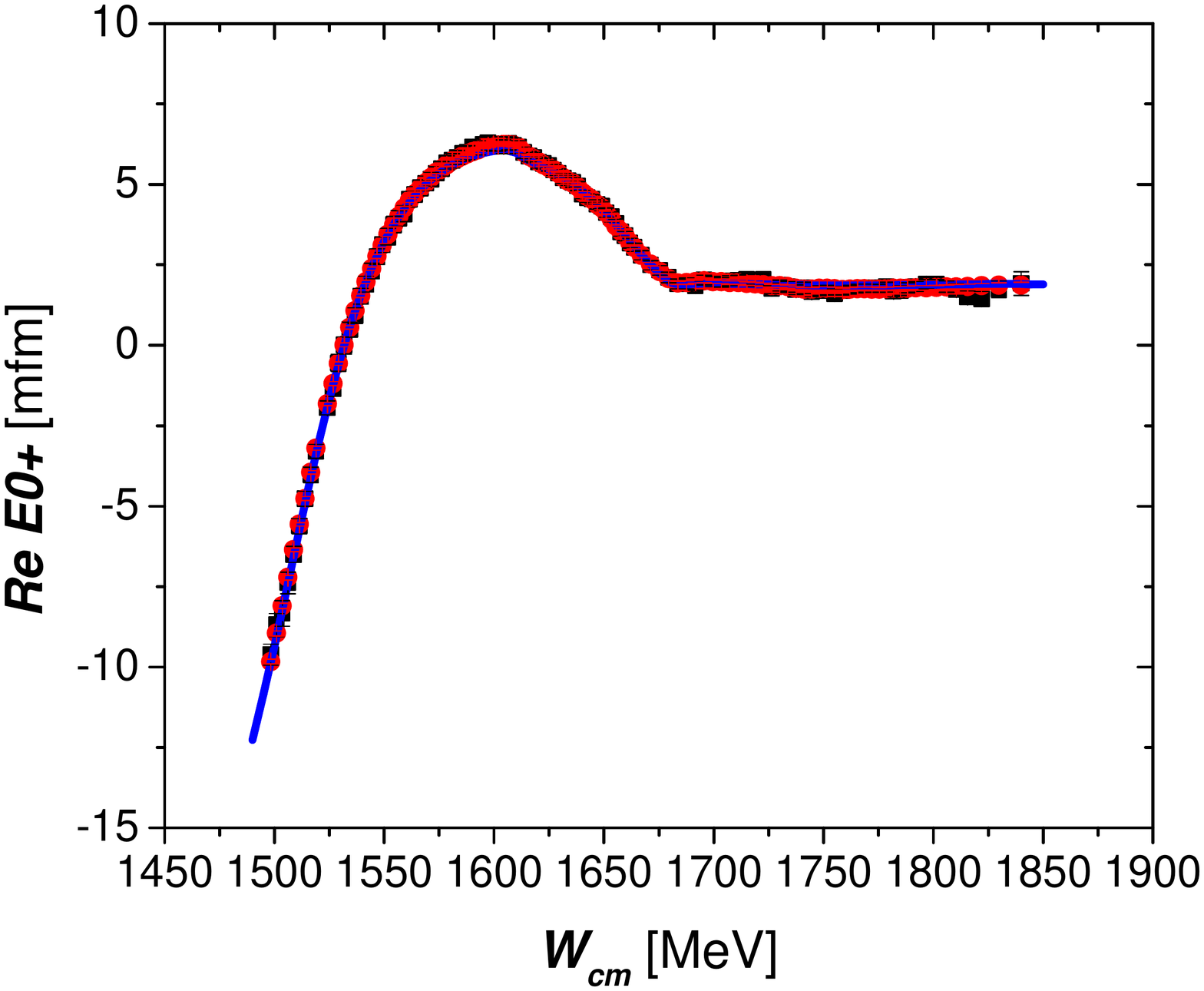}
\includegraphics[height=0.25\textwidth,width=0.4\textwidth]{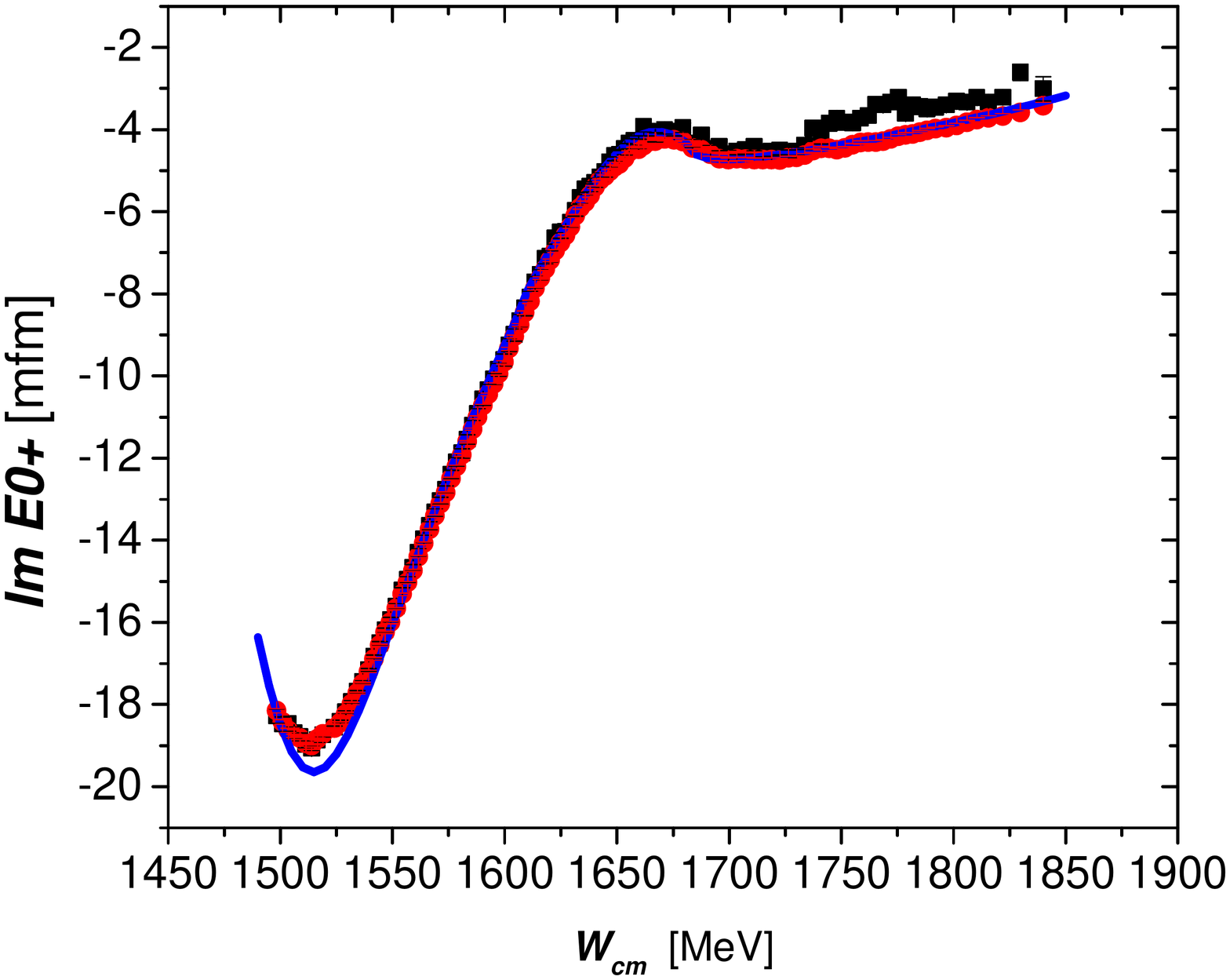} \\
\includegraphics[height=0.25\textwidth,width=0.4\textwidth]{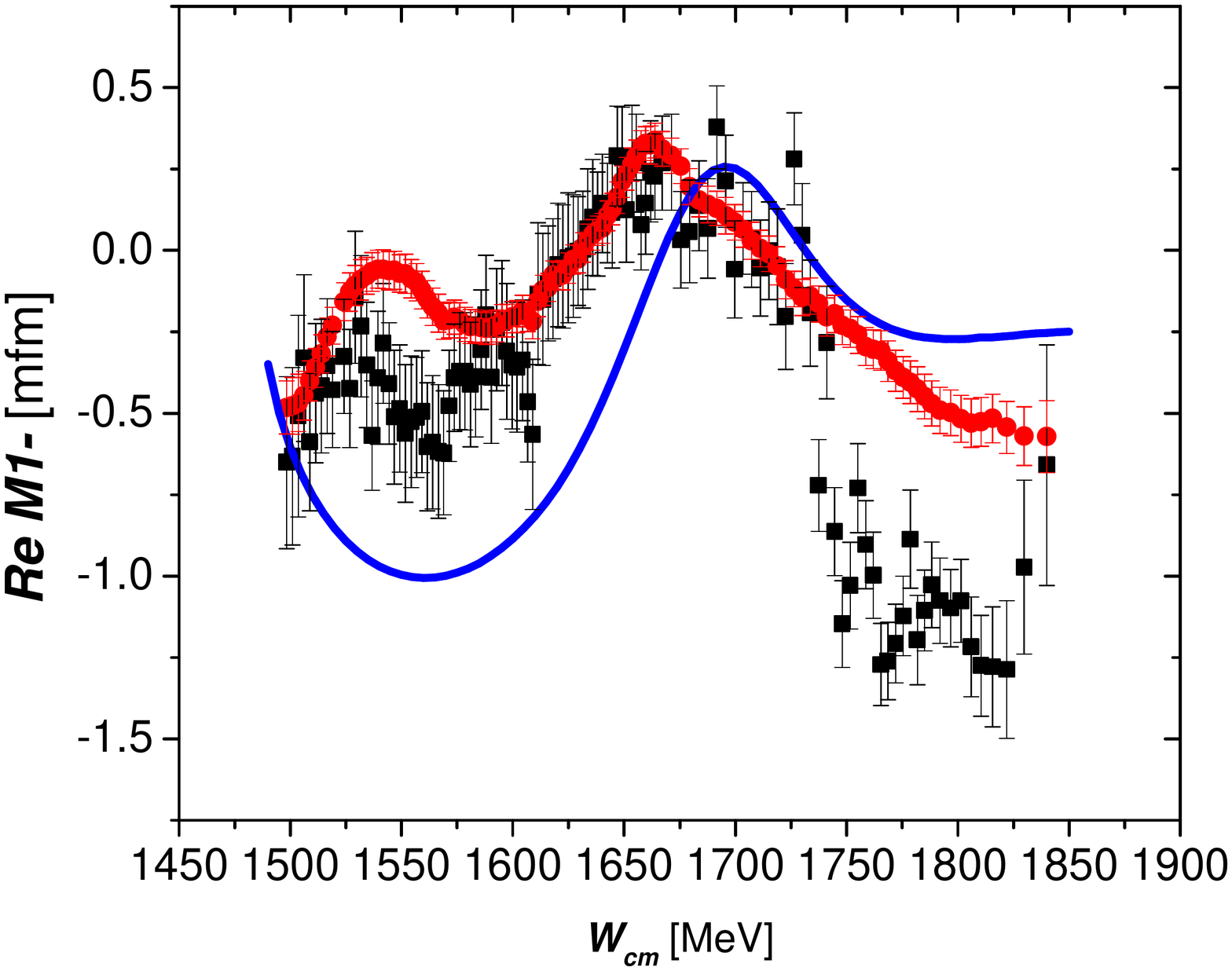}
\includegraphics[height=0.25\textwidth,width=0.4\textwidth]{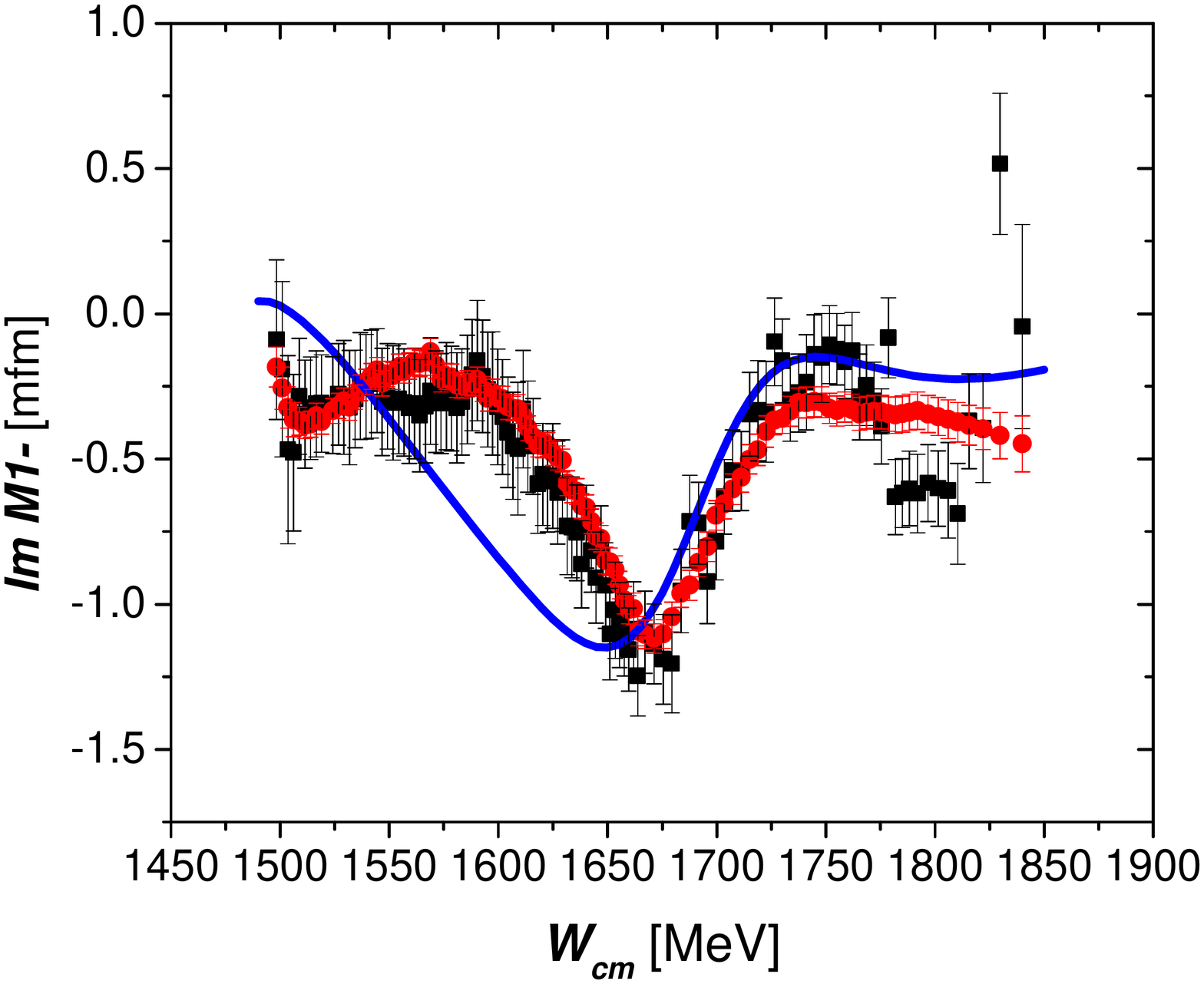} \\
\includegraphics[height=0.25\textwidth,width=0.4\textwidth]{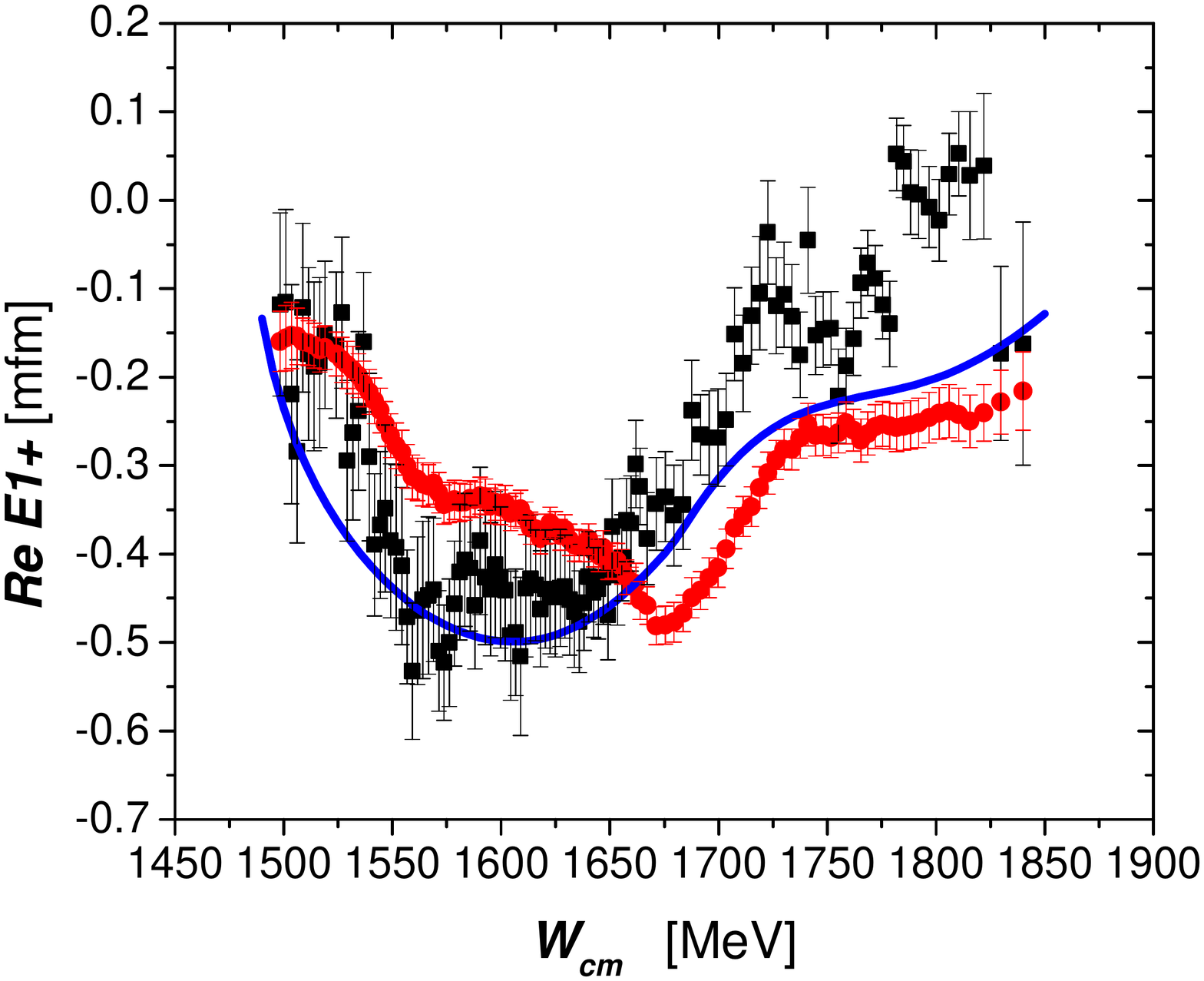}
\includegraphics[height=0.25\textwidth,width=0.4\textwidth]{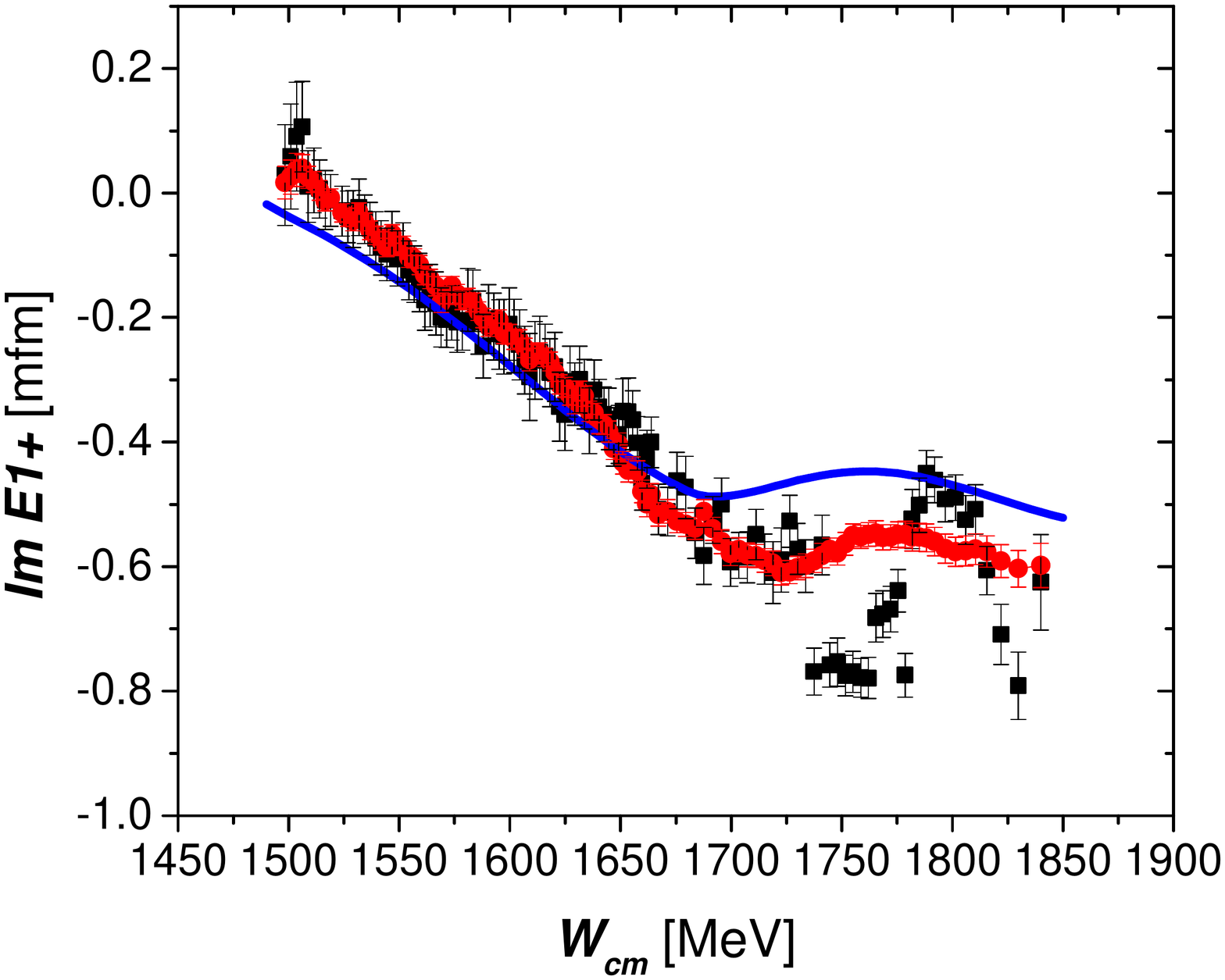} \\
\includegraphics[height=0.25\textwidth,width=0.4\textwidth]{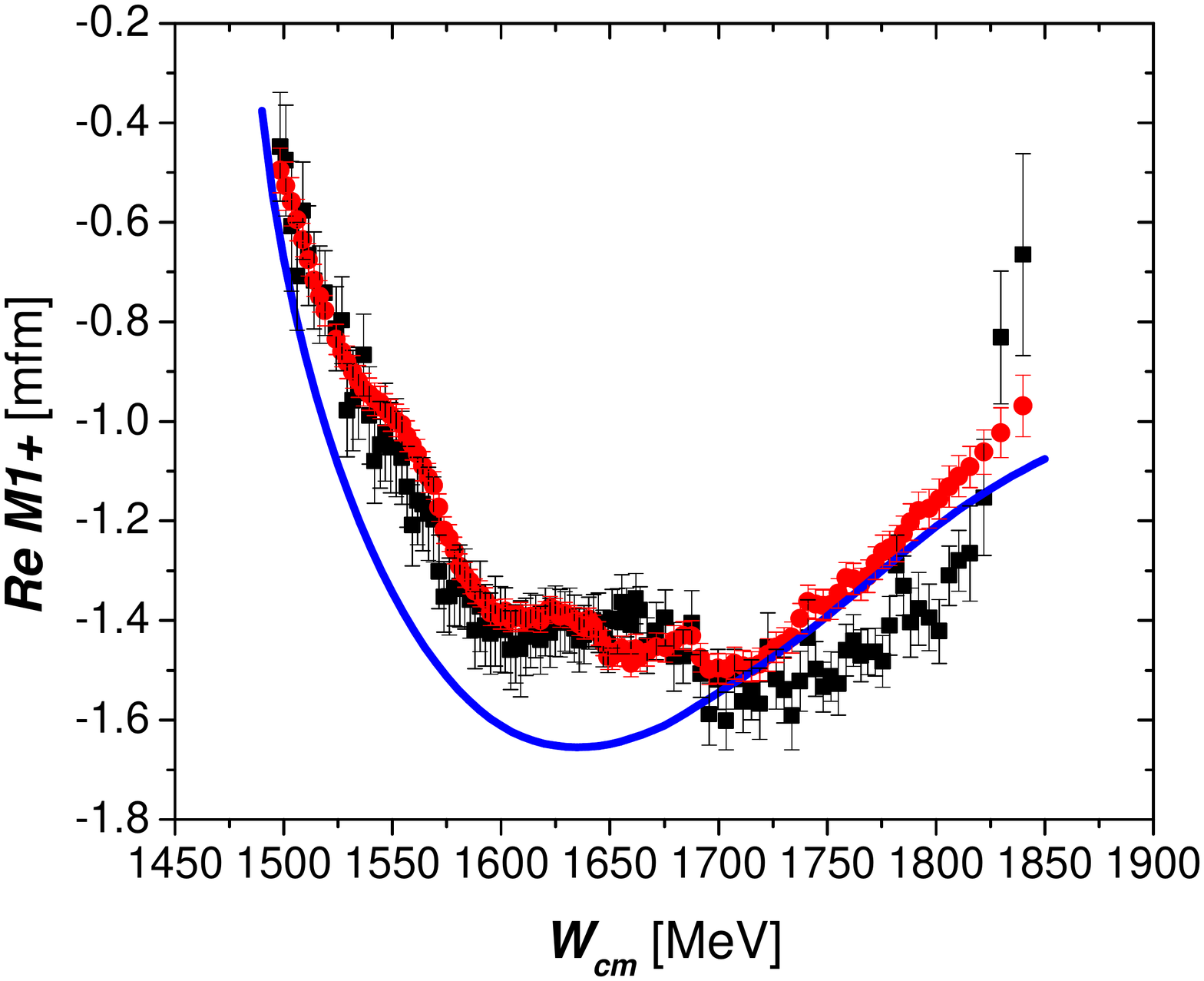}
\includegraphics[height=0.25\textwidth,width=0.4\textwidth]{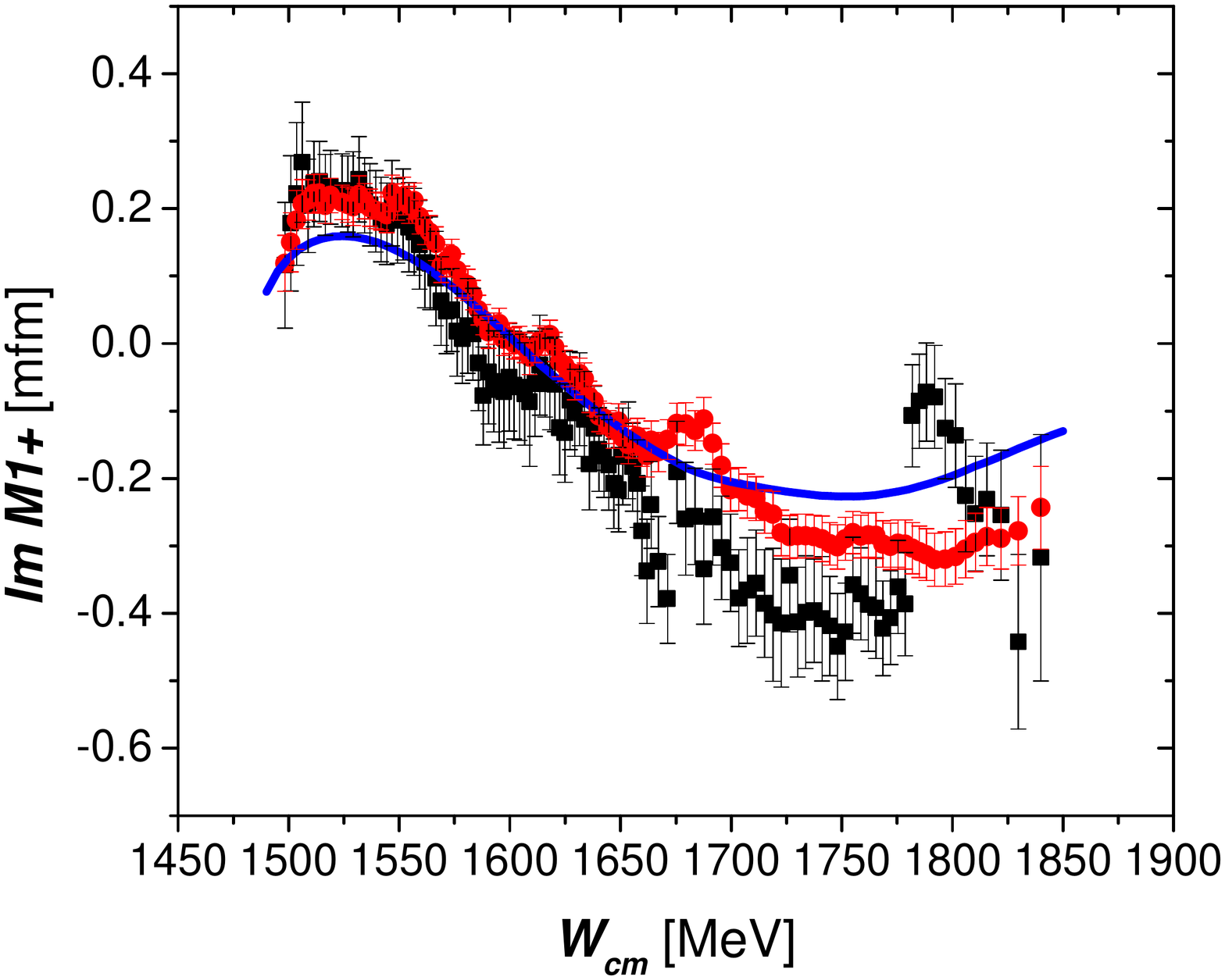} \\
\includegraphics[height=0.25\textwidth,width=0.4\textwidth]{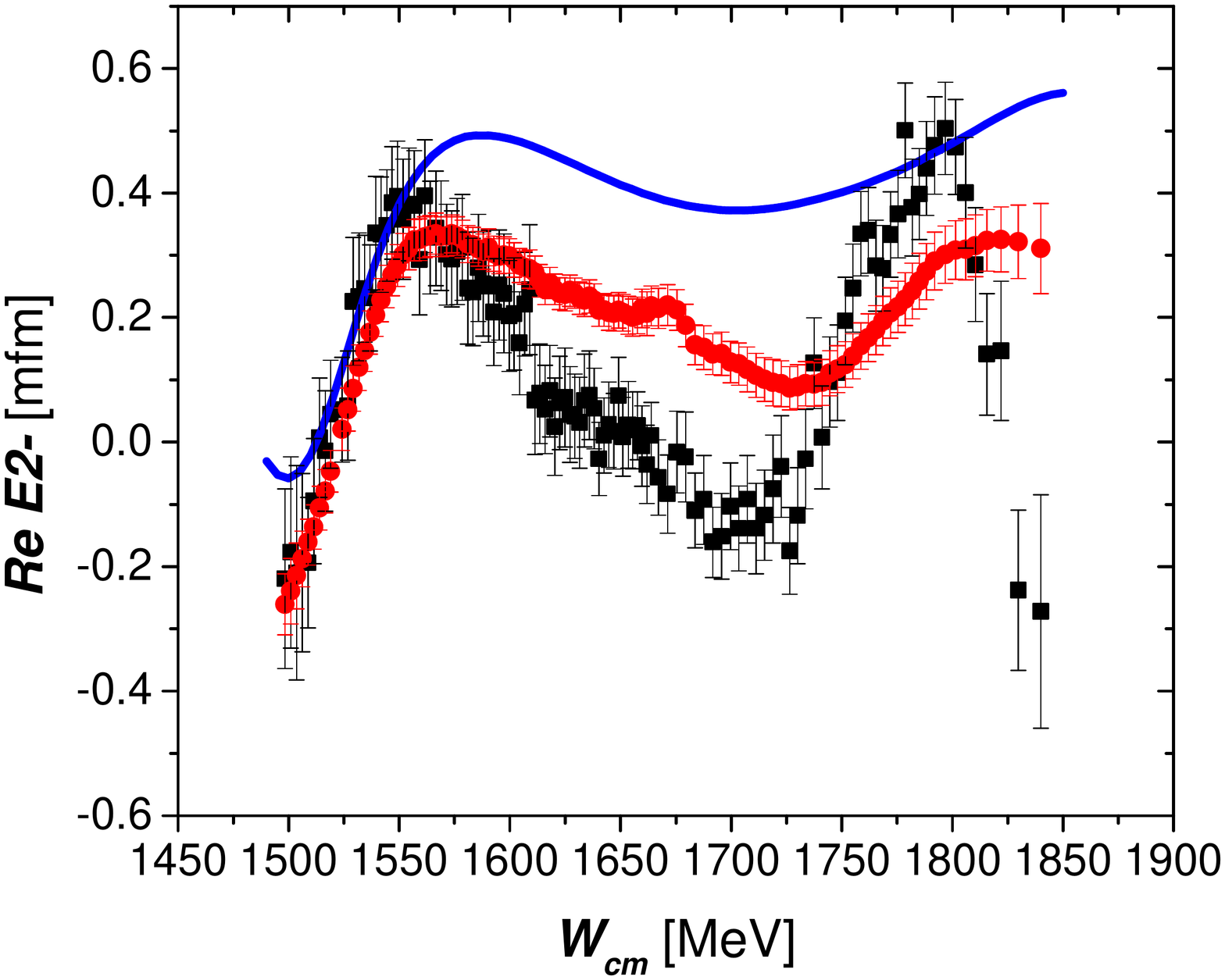}
\includegraphics[height=0.25\textwidth,width=0.4\textwidth]{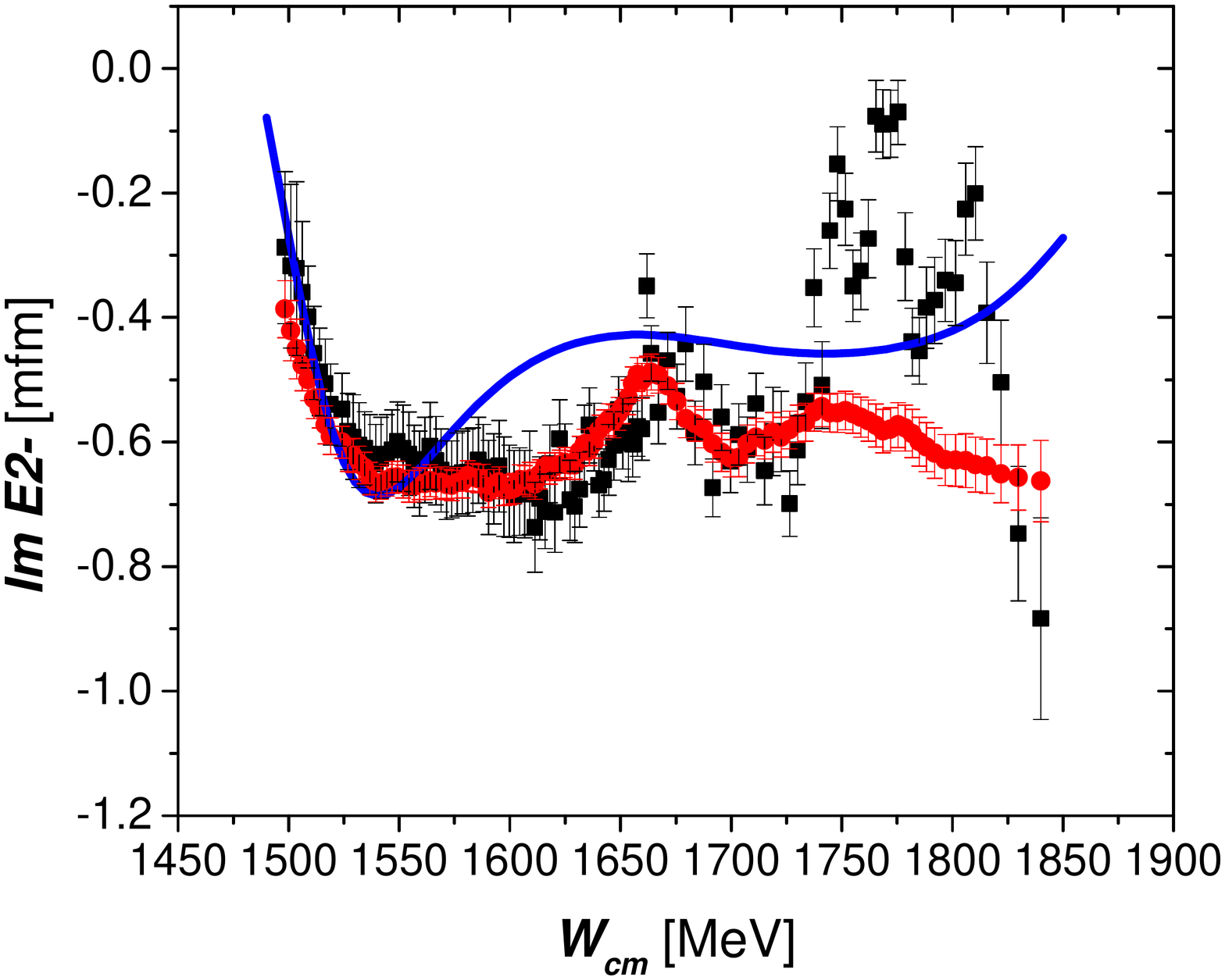} \\
\caption{\label{Fig1}(Color online) Comparison of lowest multipoles for Sol 1 \cite{Svarc2020} (black squares), Sol 1/21 (red full circles) this publication, and BG 2014-2 ED model \cite{BoGa} (full blue line) .    }
\ec
\end{figure}

\begin{figure}[h!]
\bc
\includegraphics[height=0.25\textwidth,width=0.4\textwidth]{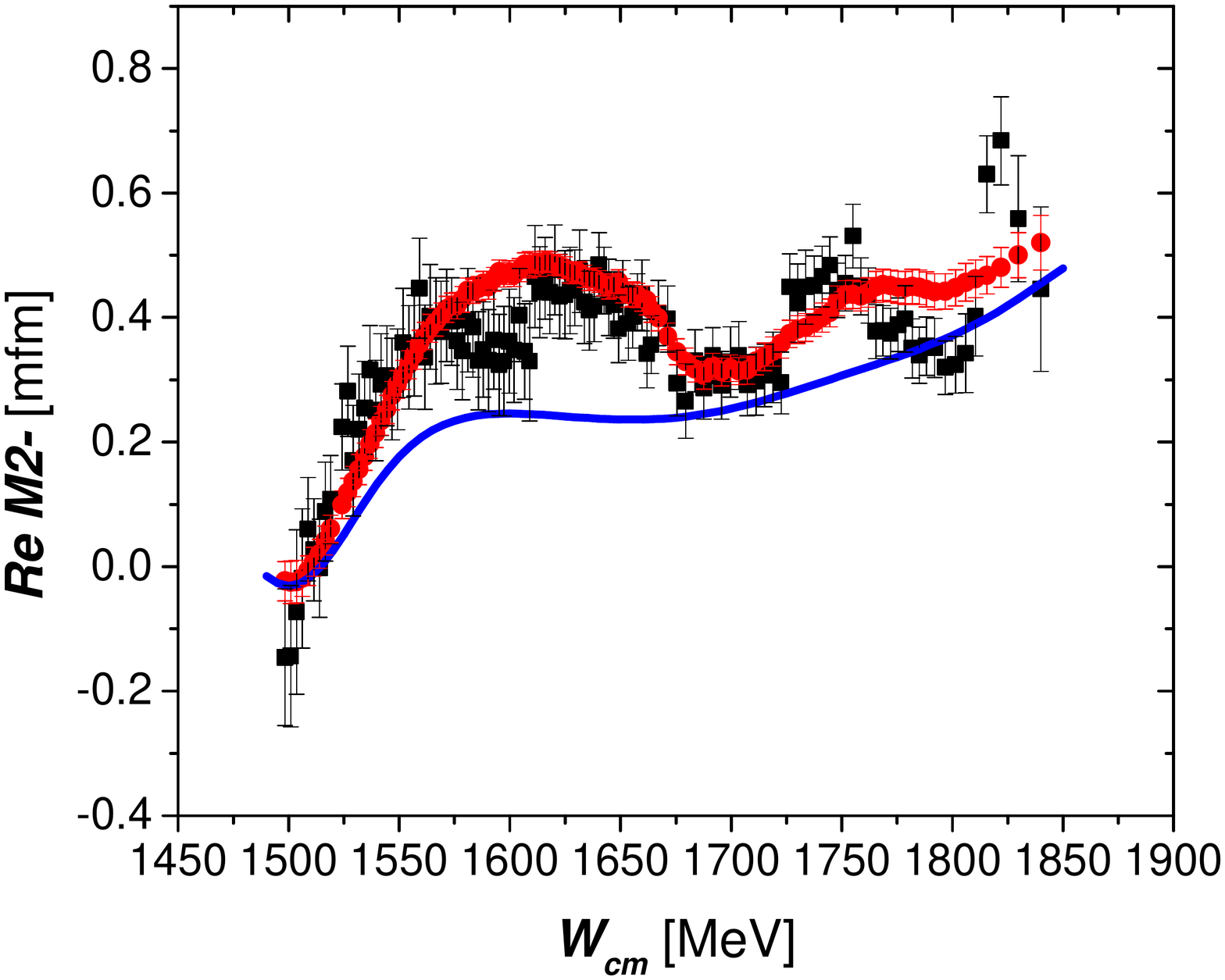}
\includegraphics[height=0.25\textwidth,width=0.4\textwidth]{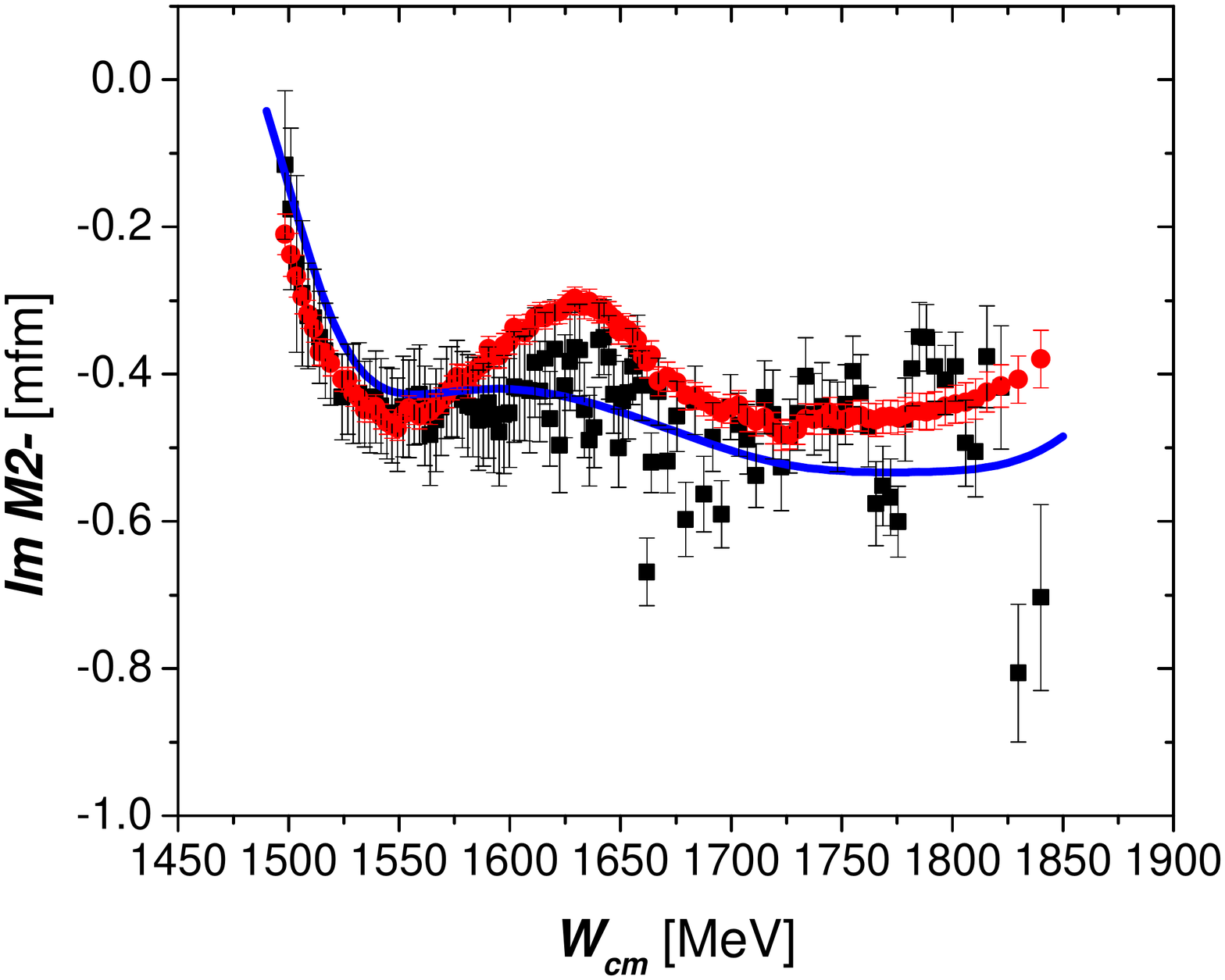} \\
\includegraphics[height=0.25\textwidth,width=0.4\textwidth]{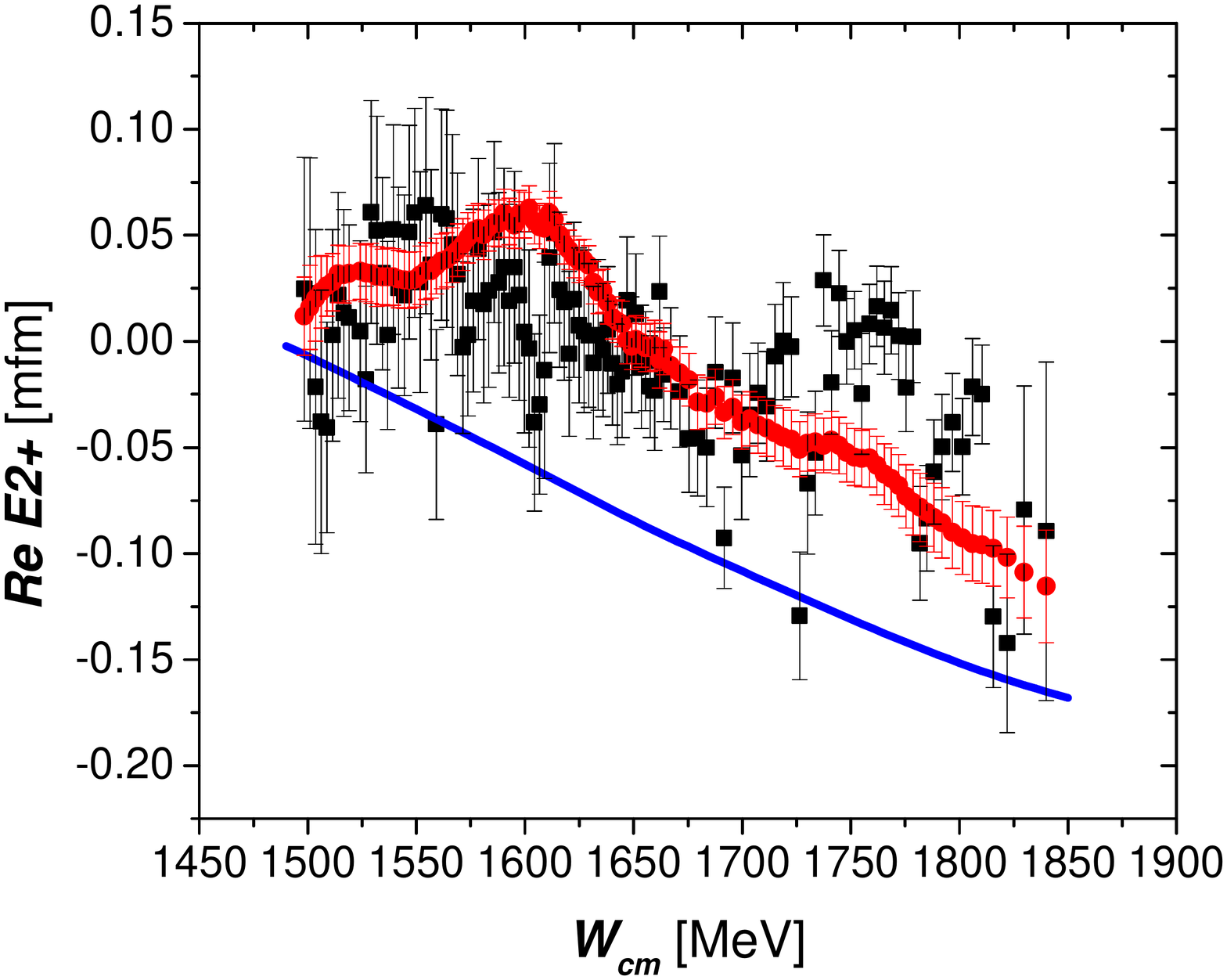}
\includegraphics[height=0.25\textwidth,width=0.4\textwidth]{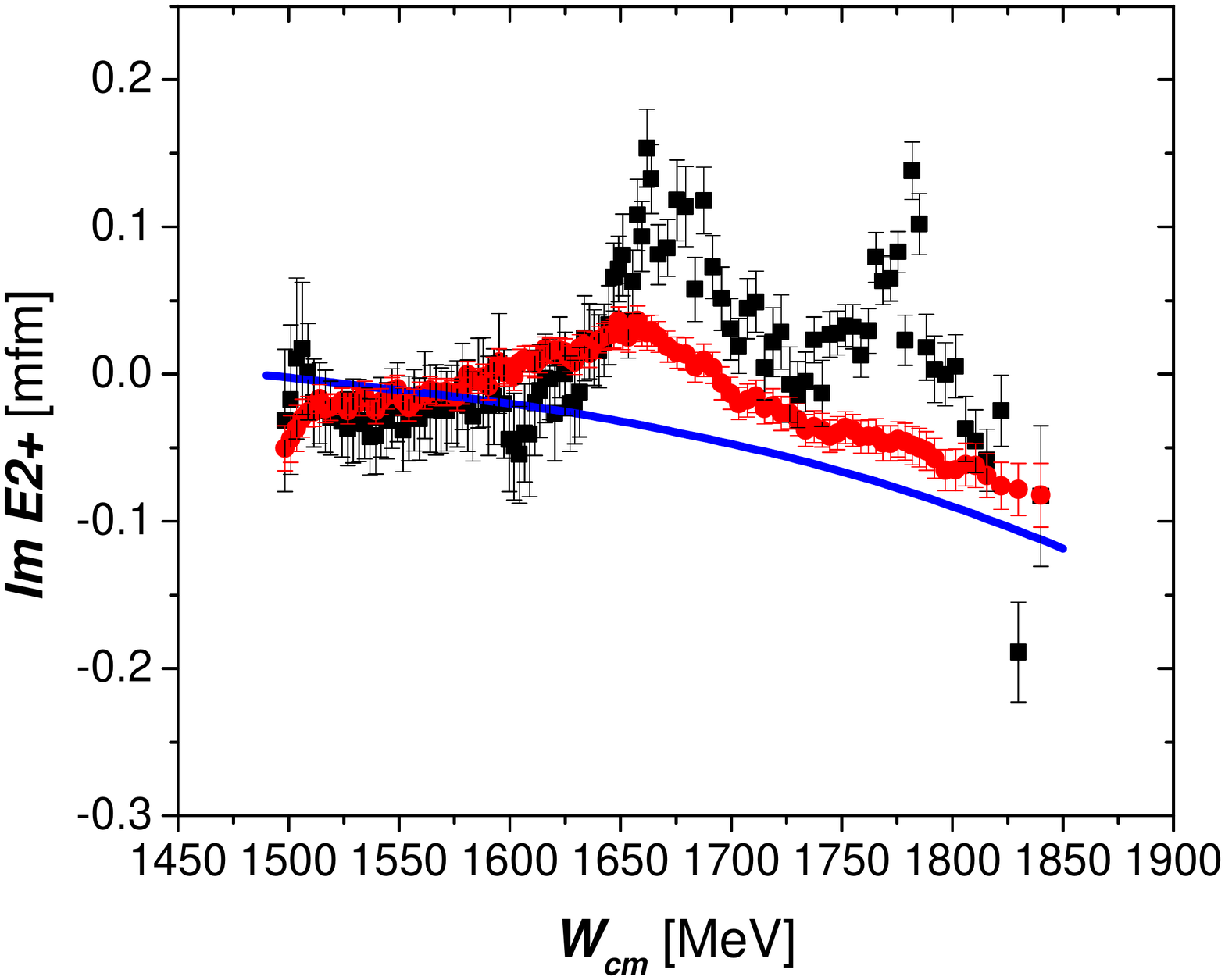} \\
\includegraphics[height=0.25\textwidth,width=0.4\textwidth]{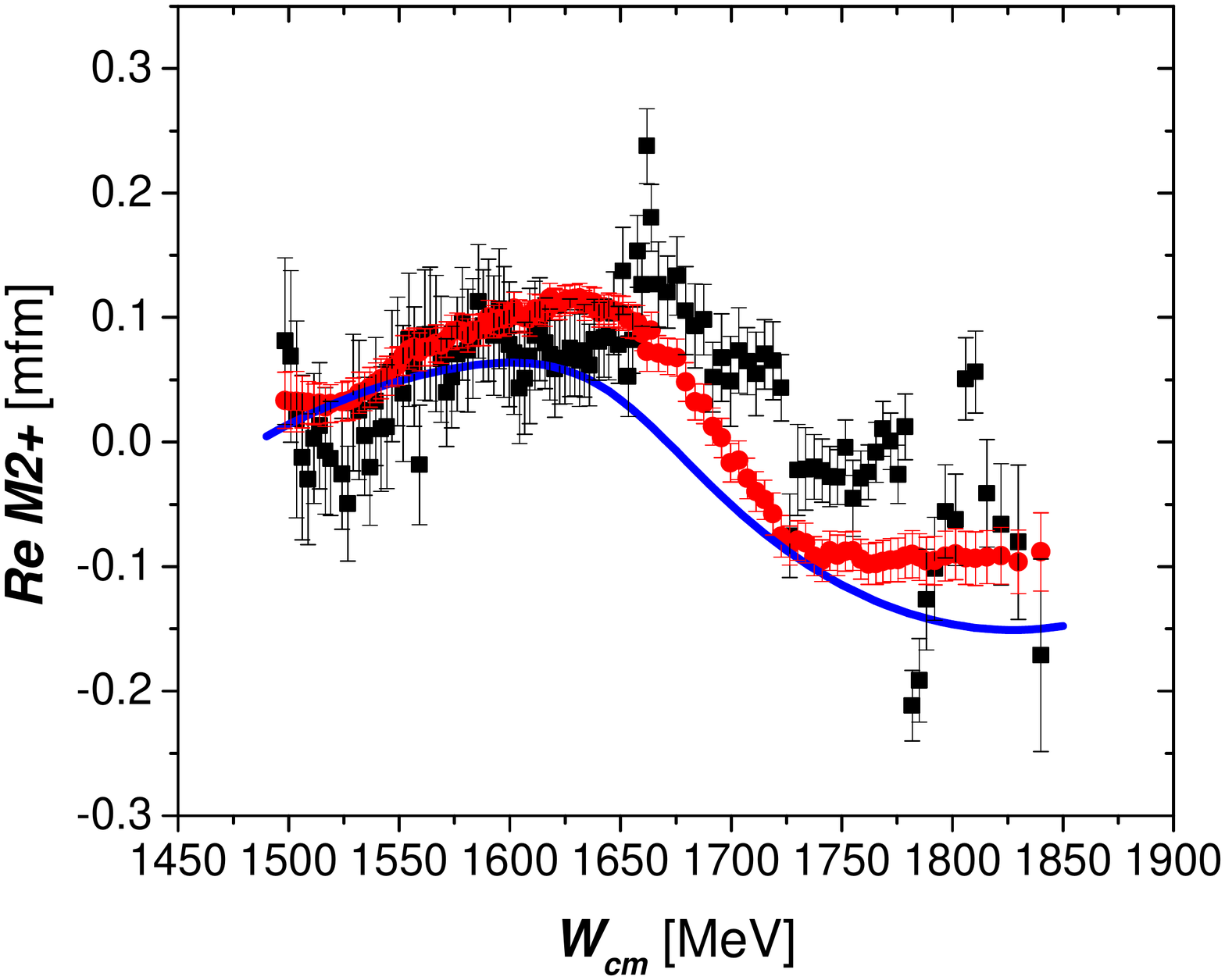}
\includegraphics[height=0.25\textwidth,width=0.4\textwidth]{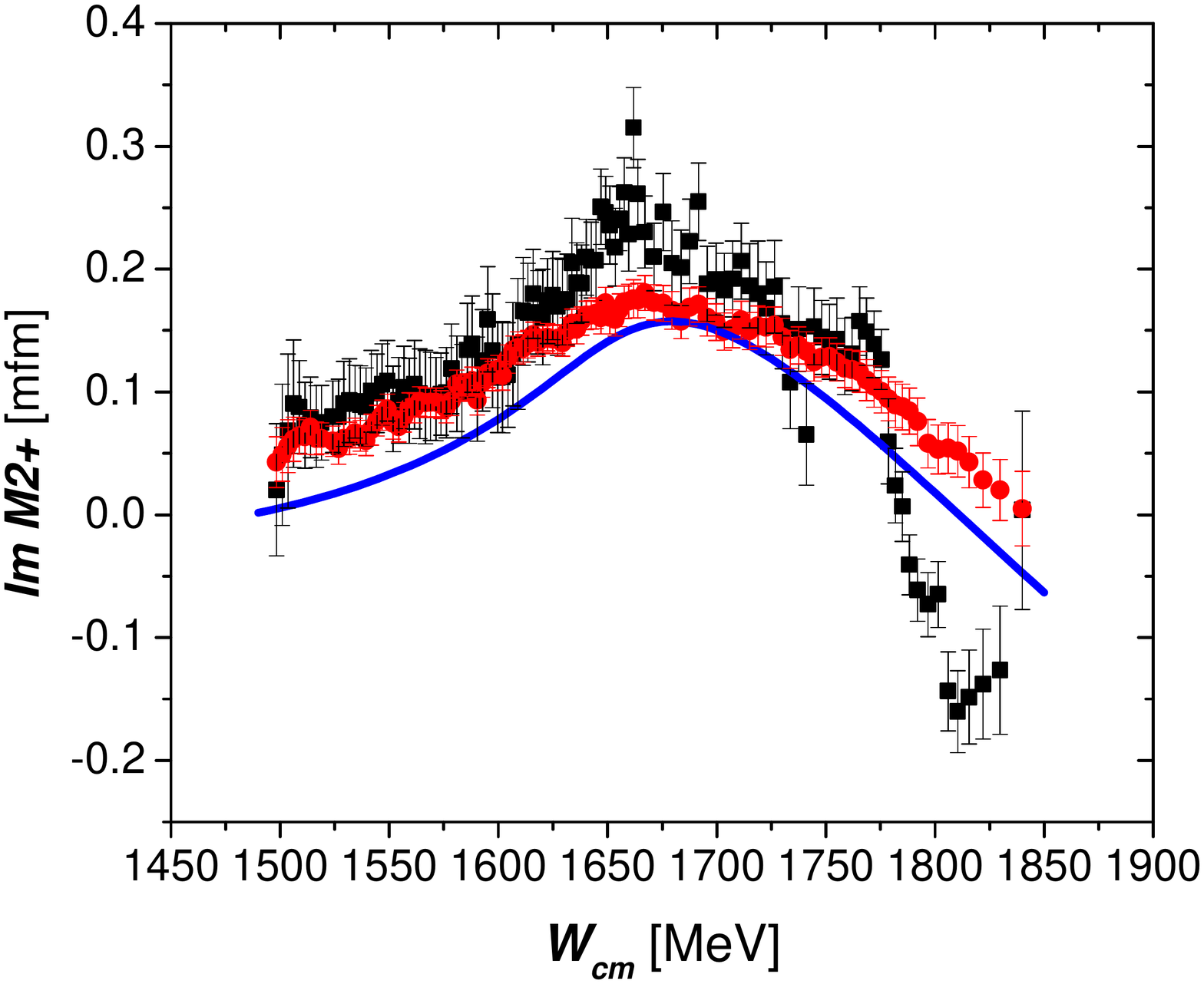} \\
\includegraphics[height=0.25\textwidth,width=0.4\textwidth]{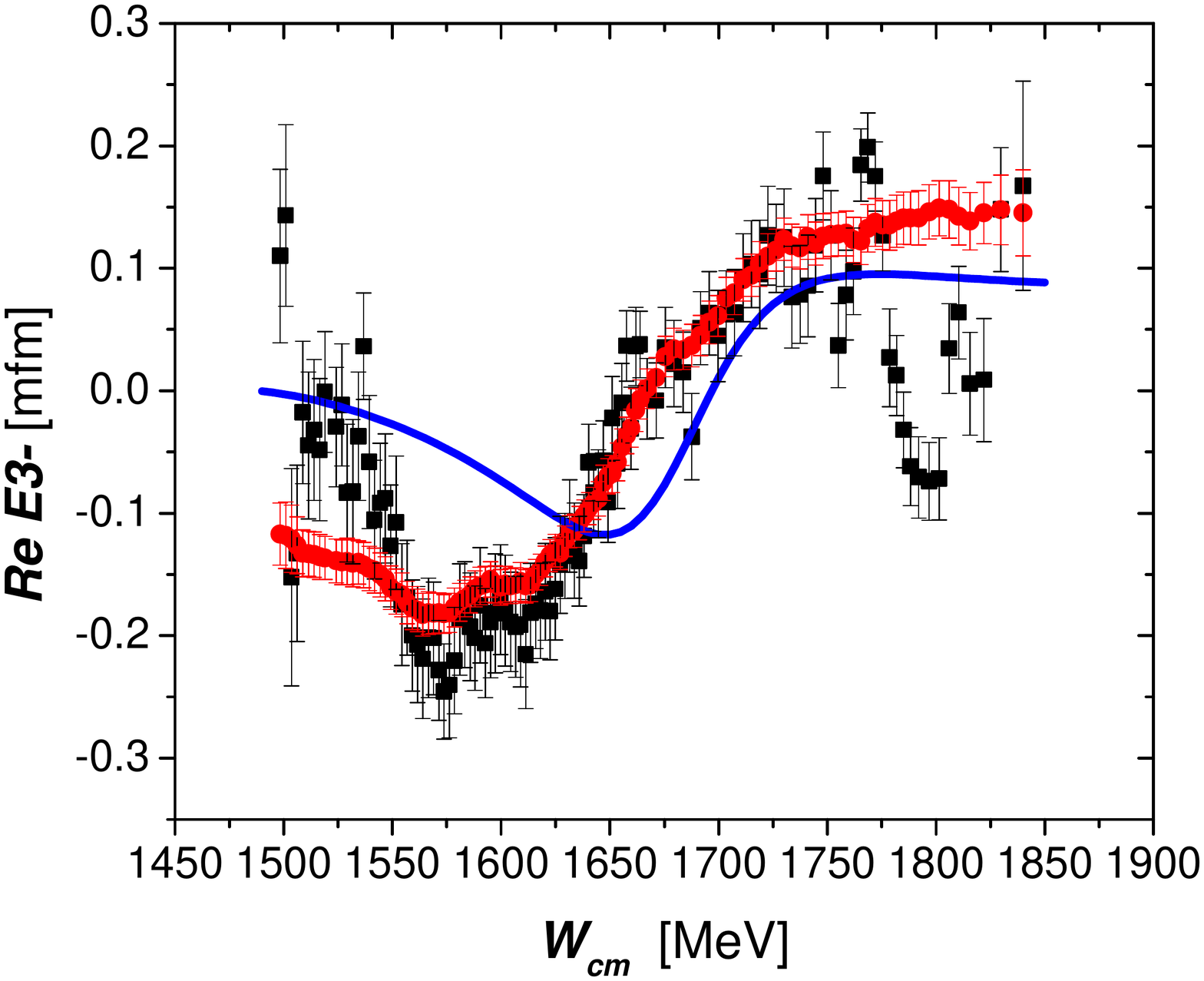}
\includegraphics[height=0.25\textwidth,width=0.4\textwidth]{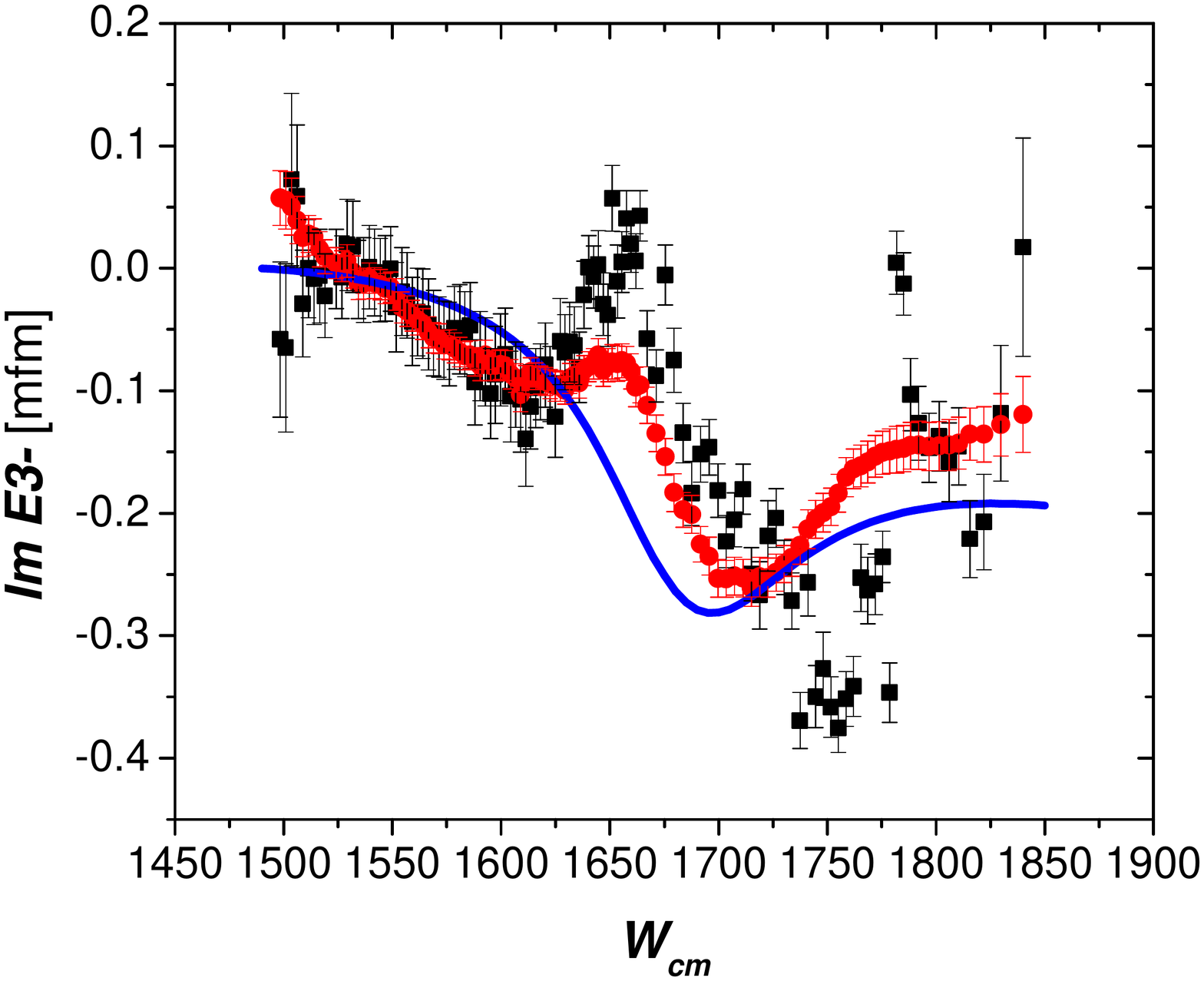} \\
\includegraphics[height=0.25\textwidth,width=0.4\textwidth]{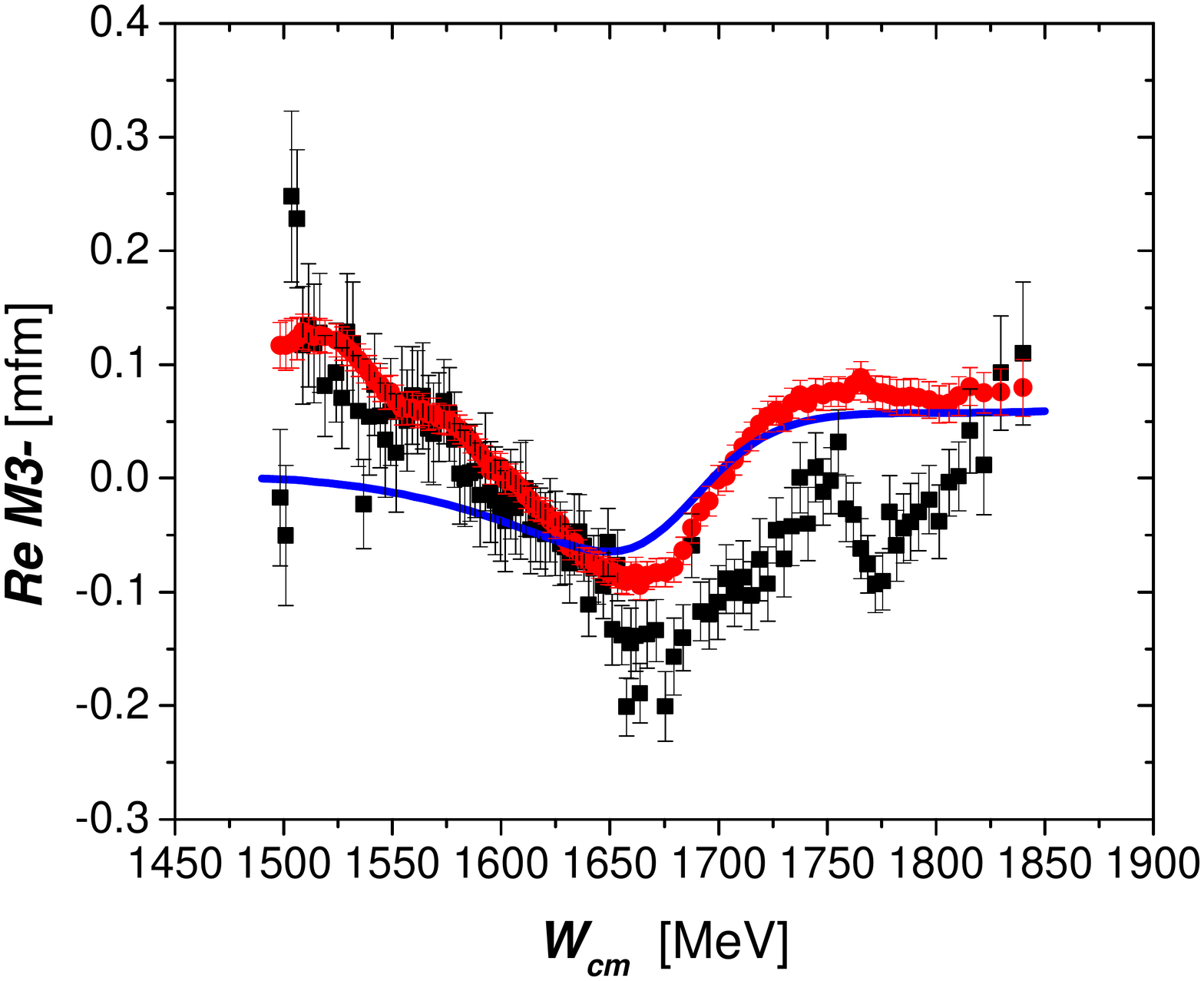}
\includegraphics[height=0.25\textwidth,width=0.4\textwidth]{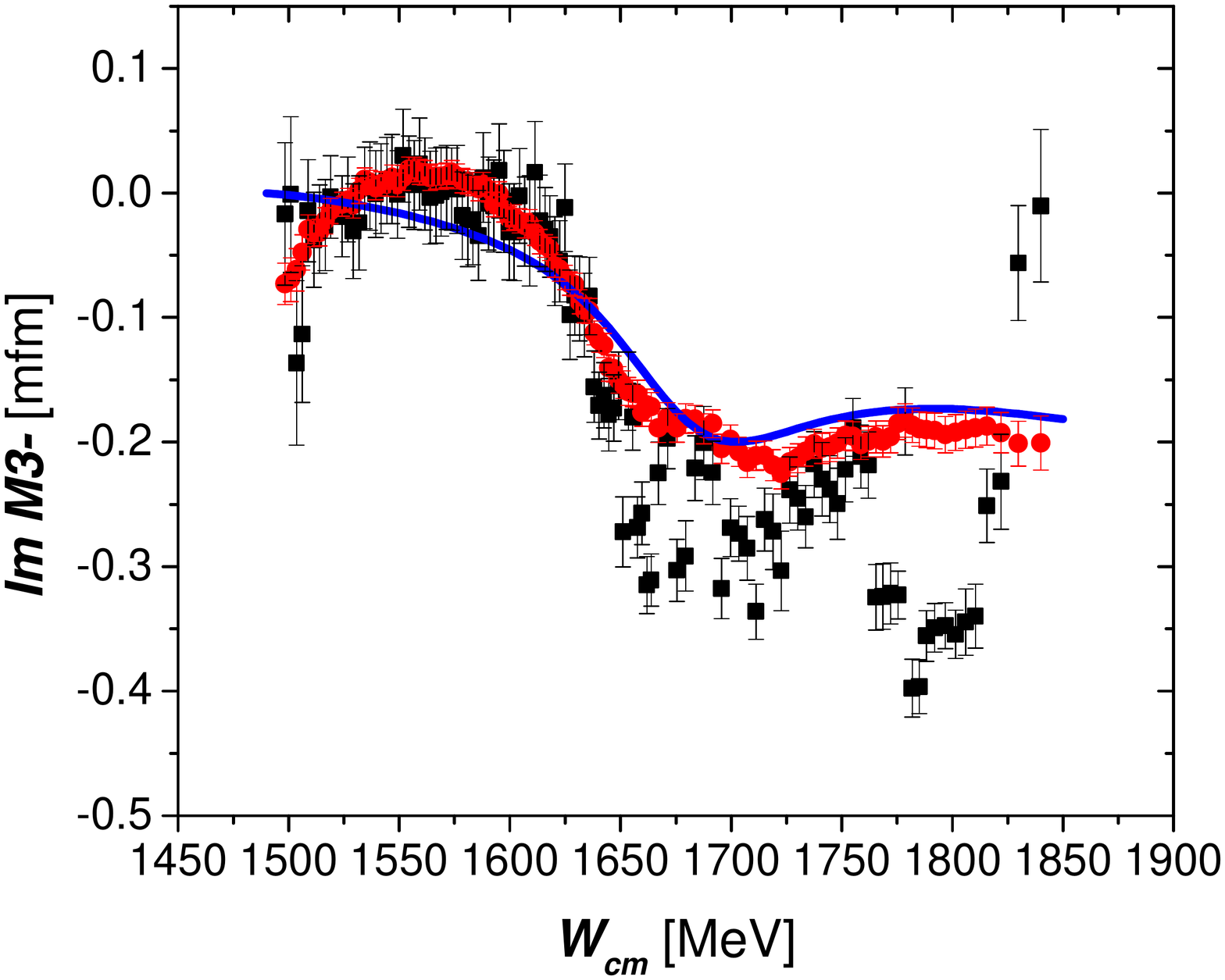} \\
\caption{\label{Fig2}(Color online) Continuation of Fig.\ref{Fig1}    }
\ec
\end{figure}

\begin{figure}[h!]
\bc
\includegraphics[height=0.25\textwidth,width=0.4\textwidth]{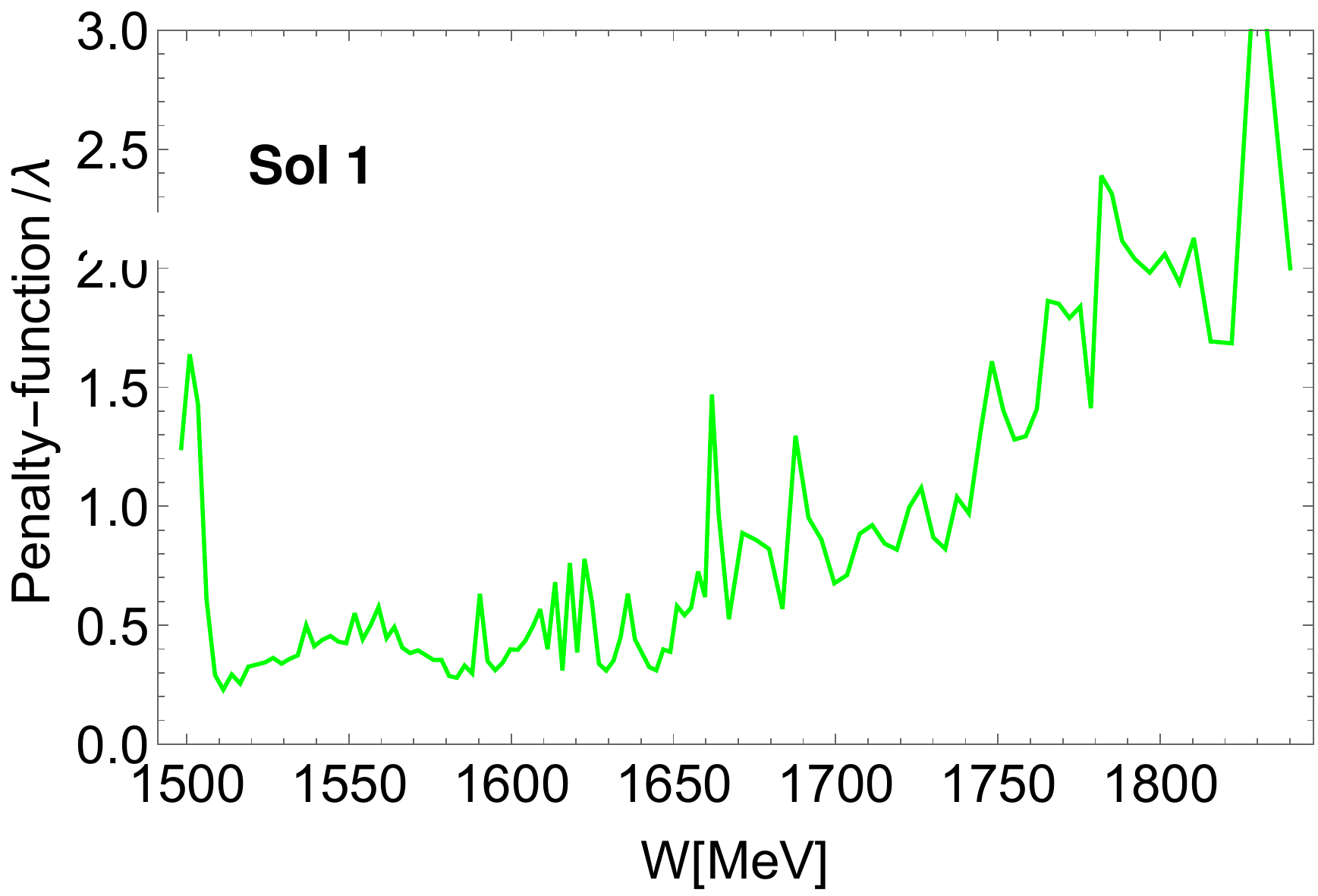} \hspace*{0.3cm}
\includegraphics[height=0.25\textwidth,width=0.4\textwidth]{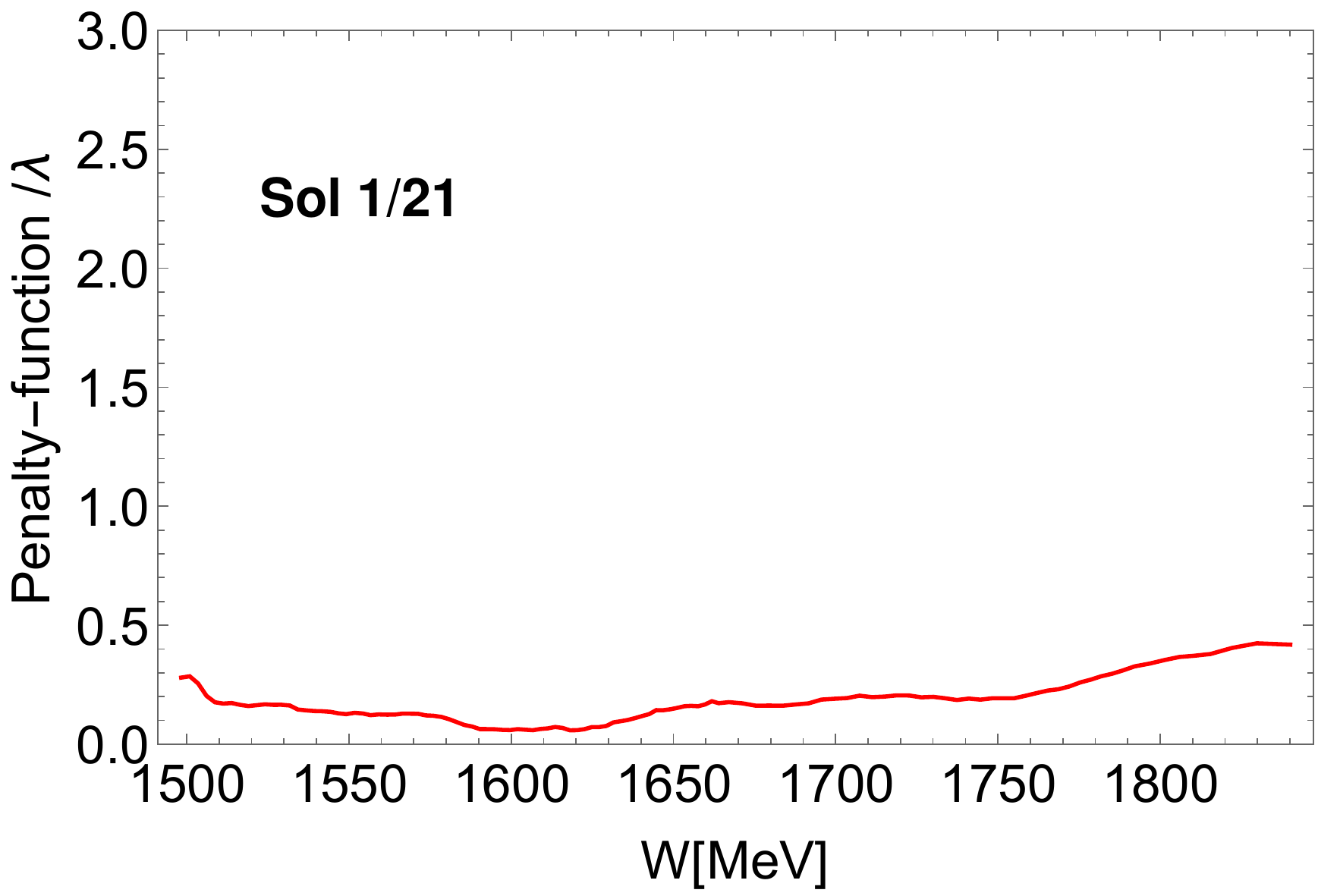} \\
\caption{\label{Fig3}(Color online) Comparison of penalty function $\cal{P}$ for Sol 1 and Sol 1/21.    }
\ec
\end{figure}

From our earlier research \cite{Svarc2018}  we know that the natural candidate for such an effect is  the change of reaction amplitude phase. As the essential part of our AA/PWA model is the penalty factor $\cal{P}$, and the possible change of phase is introduced via this factor, we shall put this factor under magnifying glass. This factor introduces two phenomena: ensures the continuity in partial waves, and ensures the proximity of phase to the known input phase of the ED model which, I repeat, is a matter of our choice. In ref~\cite{Svarc2018} we have opted for the moderate constraint, and used $\lambda_{pen}=10$. To fully test the consequence of such a choice, we have  generated a new solution with much stronger constraint using  $\lambda_{pen}=500$  called Sol 1/21. We show the outcome in Figs.(\ref{Fig1}) and (\ref{Fig2}). To quantify the amount of likeness of our new solution phase with original phase, we also show the penalty function  $\cal{P}$ for both solutions in Fig.~(\ref{Fig3}).

\begin{figure}[h!]
\bc
\includegraphics[height=0.25\textwidth,width=0.42\textwidth]{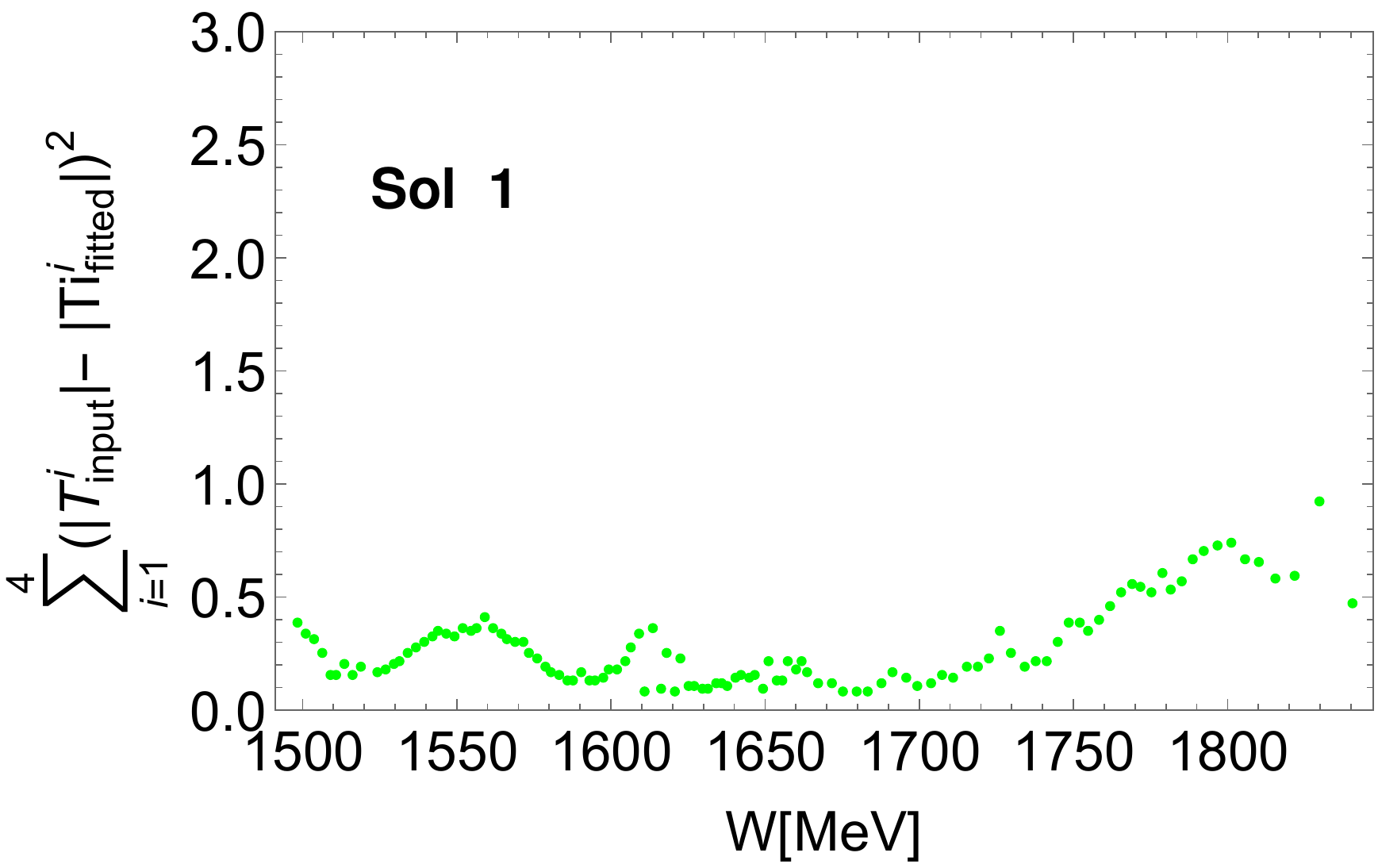}
\includegraphics[height=0.25\textwidth,width=0.42\textwidth]{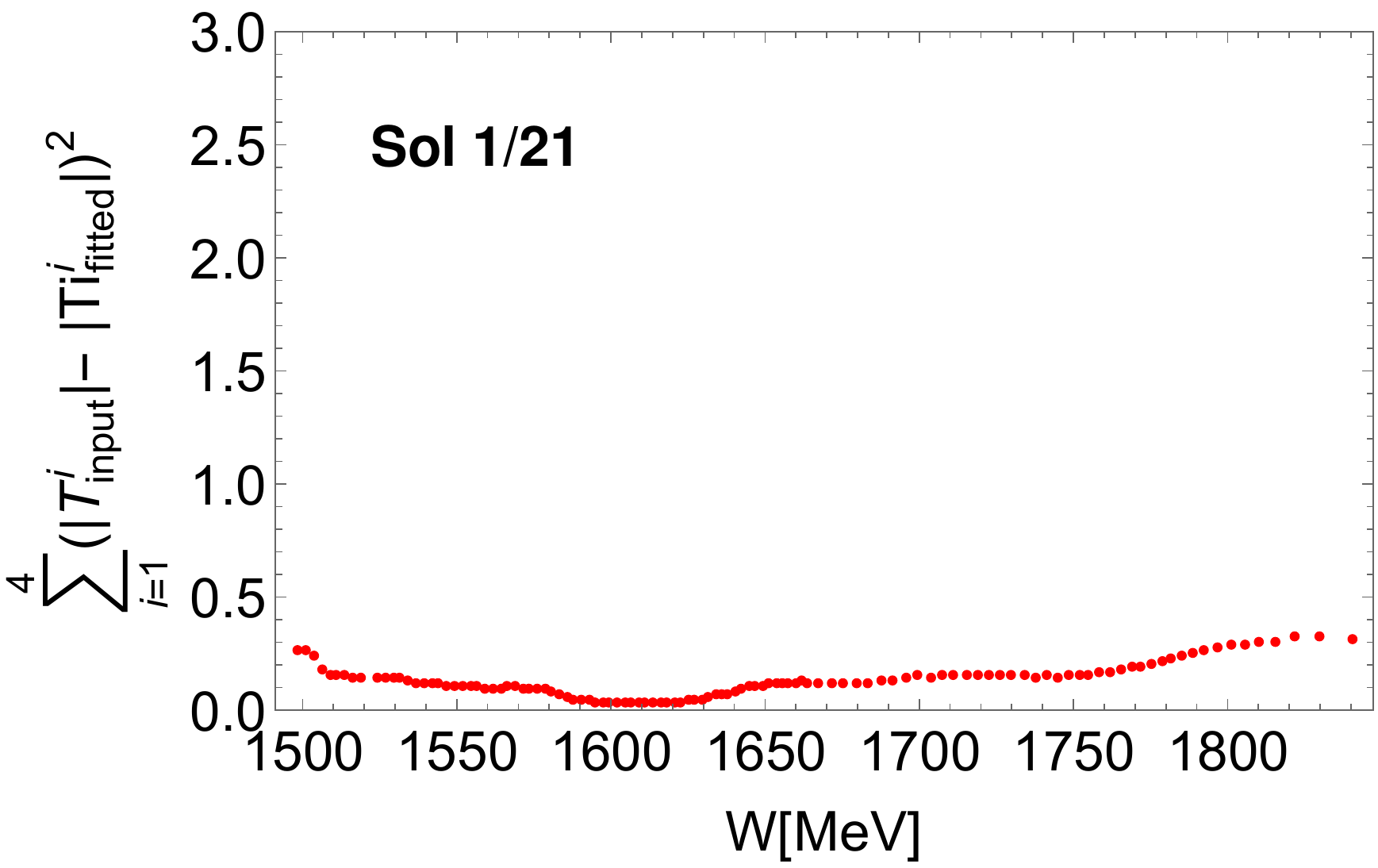} \\
\caption{\label{Fig4}(Color online) Comparison of absolute values of penalty function $\cal{P}$ for Sol 1 and Sol 1/21.    }
\ec
\end{figure}

\begin{figure}[h!]
\bc
\includegraphics[height=0.25\textwidth,width=0.4\textwidth]{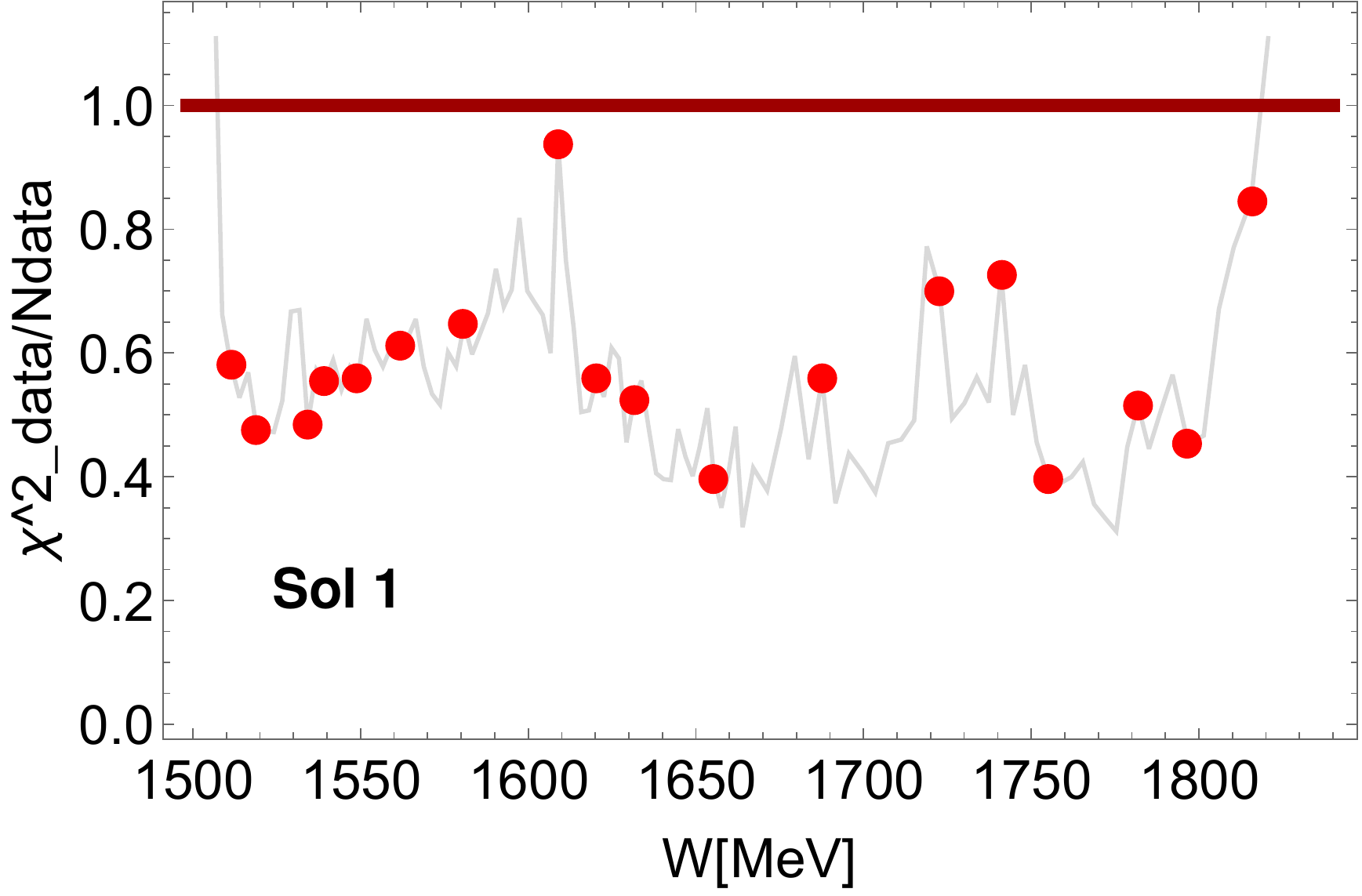} \hspace*{0.3cm}
\includegraphics[height=0.25\textwidth,width=0.4\textwidth]{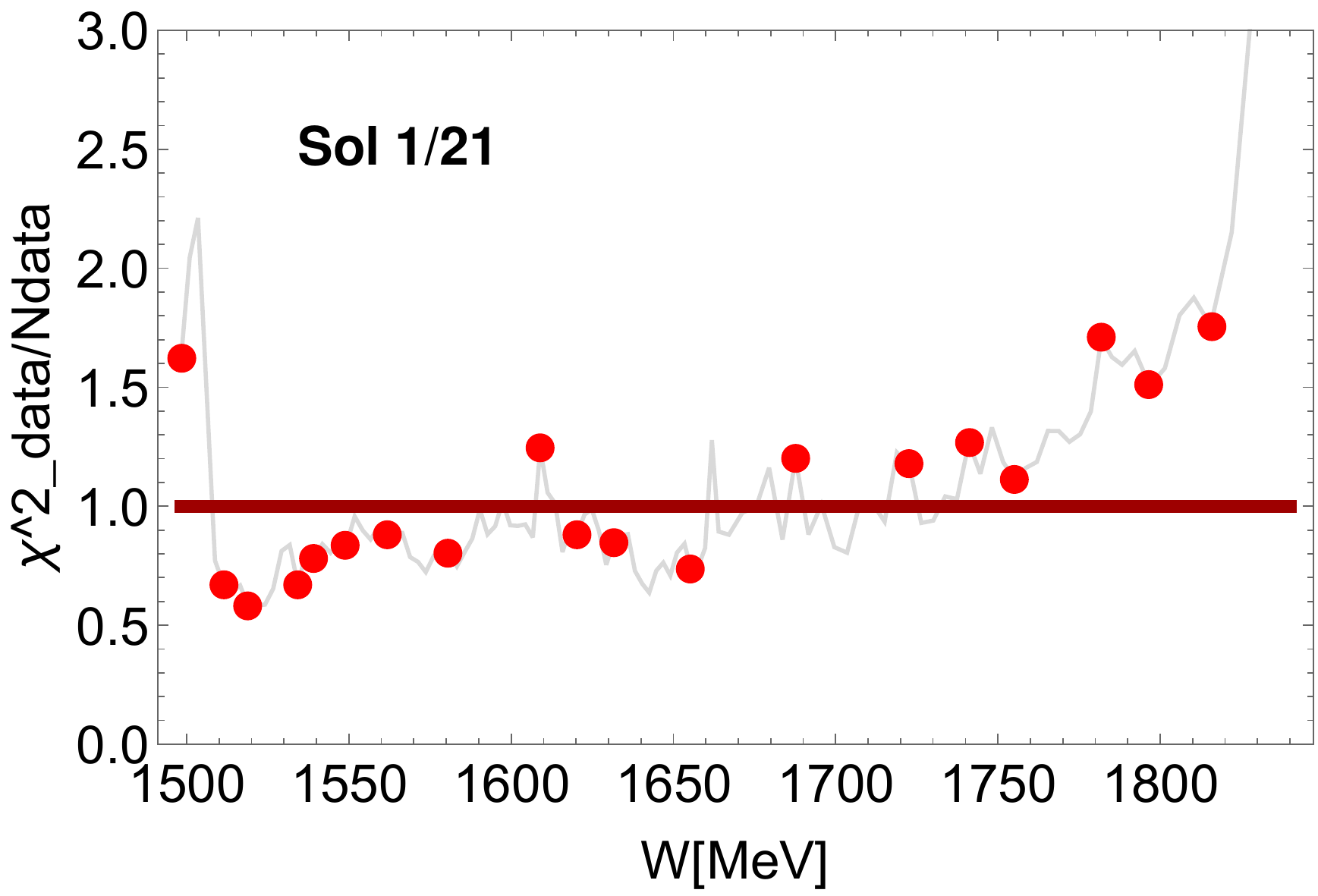} \\
\caption{\label{Fig5}(Color online) Comparison $\chi^2/data$ for Sol 1 and Sol 1/21.    }
\ec
\end{figure}

First form Figs.~(\ref{Fig1}) and (\ref{Fig2}) we see that new multipoles of Sol 1/21 (red symbols) are notably different from the old ones of Sol 1 (black squares), and that they are much smoother. From Fig.~(\ref{Fig3})   we see that the normalized penalty function of Sol 1/21 (penalty function divided by the penalty coefficient~$\lambda_{pen}$), as expected, is much lower that the normalized penalty function of Sol 1, and uniform over energies. Let us stress that it should not vanish, as transversity amplitudes generated in \emph{Step 1} are in principle discontinuous both in energy and angle, and the continuity of the solution is imposed by the very weak condition of constraining the fit to the neighbouring energy solution. For Sol 1, deviation of penalty function from the constraining value is also notably rising with energy. From this figure alone it is not clear whether this increase is due to change of phase or to the change of absolute value as the penalty function $\cal{P}$  is defined on full amplitudes. Therefore, we in Fig.(\ref{Fig4}) show only the difference of absolute values of Sol 1 and Sol 1/21. We see that these quantities are very stable in energy, and both of the same order of magnitude. This shows that increase of penalty function of Sol 1 is solely due to the change of phase. In addition, let us illustrate another claim made before: strengthening  the constraint (imposing the phase closer to the ED phase) spoils the agreement with the data as the phase of ED solution is not ideal for  phase dependent observables.  Therefore, in Fig.(\ref{Fig5}) we show $\chi^2/data$ (chi**2 per data point) for both solutions. The $\chi^2/data$ for Sol 21/1 is still good, but the $\chi^2/data$ for Sol 1 is notably better.

\clearpage

\clearpage

\clearpage

\newpage \noindent
Let us summarize the results of L+P analysis  shown in  Table~\ref{Table1}.
\\ \\ \indent
As said before, poles of the energy dependent solution BG2014-2 correspond to the values published by Bonn-Gatchina group \cite{BoGa}. Extraction of the first pole in the S-wave is unreliable as the pole lies in the vicinity of $\eta$-photoproduction threshold. This means that it lies on the different Riemann sheet, so the residue value has to be big to have any influence upon the data. However, L+P fit unquestionably needs this pole, but the residues are big, and unreliable meaning that we can generate similar solutions with a wide choice of different backgrounds. This confirms a general problem of residue in L+P formalism. As pole-background separation in Laurent expansion is not theoretically well defined, we rely on a fit, so unquestionably we can have different backgrounds which differently combine with the residue of the singular part giving identical results. And this is the  general feature of all fits. Anyway, the agreement with known values is good.
\\ \\ \indent
Next we summarize the analysis of SE multipoles obtained with AA/PWA method. First let us stress that the poles of SE solutions should be close to BG2014-2 values, but definitely should not be identical to them. Namely, ED BG2014-2 is coupled-channel model, so it is expected that the pole parameters will be formed as the overall agreement of the fit to all channels. SE partial waves are on the other hand a single-channel quantity, and will be ideally matched only to this particular channel. It is theoretically known that these poles should be identical in all channels, but in practice experimental error introduces unwanted uncertainty, especially when all channels are treated simultaneously. Therefore, poles of ED models are expected to differ from poles of SE PWA. We can only expect that our results are within the confidence level of PDG. In this table we systematically see that Sol 1 poles are less certain, sometimes with unrealistic errors, and unrealistic residues ($P_{13}$  $3/2+$ partial wave), and the obtained $\chi^2$ is poor. On the other hand, Sol 1/21, the solution with the phase much closer to ED phase, has much more reliable poles. First  we find that the number of poles needed to obtain a good L+P fit, is identical to the number of 4*resonances reported in PDG\footnote{With the exception of $P_{11}$ $1/2-$ partial wave what will be discussed later.}.  Pole masses are in principle within one standard deviation with PDG results, pole width are somewhat less reliable, but still acceptably good. In addition, looking at Figs.(\ref{Fig1}) and (\ref{Fig2}) we see that Sol 1/21 solution is much smoother than Sol 1 as some structures especially at higher energies are smeared out, and much closer to theoretical BG2014-2 ED value. This is not at all surprising, and this solution is much more constrained, so has to be smoother and closer to ED value.

\subsection{New $P_{11}$  $ 1/2+$ state?}
People dislike to speculate, I do not. Therefore, I give you one speculative interpretation of effects in $P_{11}$  $ 1/2+$ state.

\begin{figure}[h!]
\bc
\includegraphics[width=0.92\textwidth]{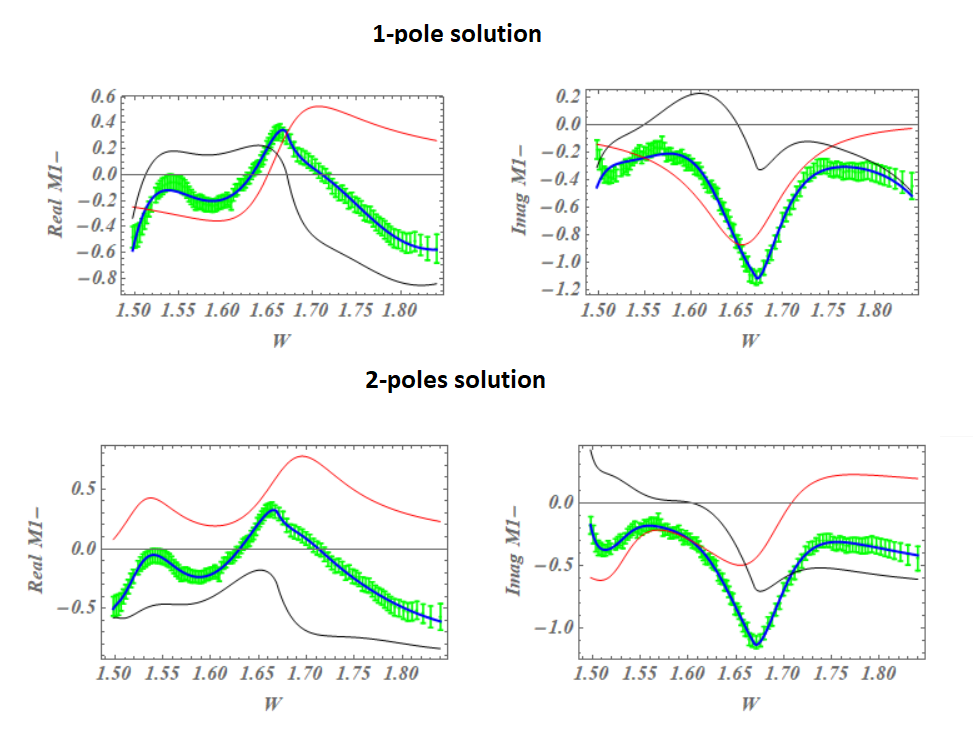}
\caption{\label{Fig6}(Color online) 1-pole and 2-pole L+P fits of  $M_1^-$ multipole. Blue lines represent the L+P fits, while red and black lines represent resonant and background contributions respectively.}
\ec
\end{figure}

\begin{figure}[t!]
\bc
\includegraphics[width=0.51\textwidth]{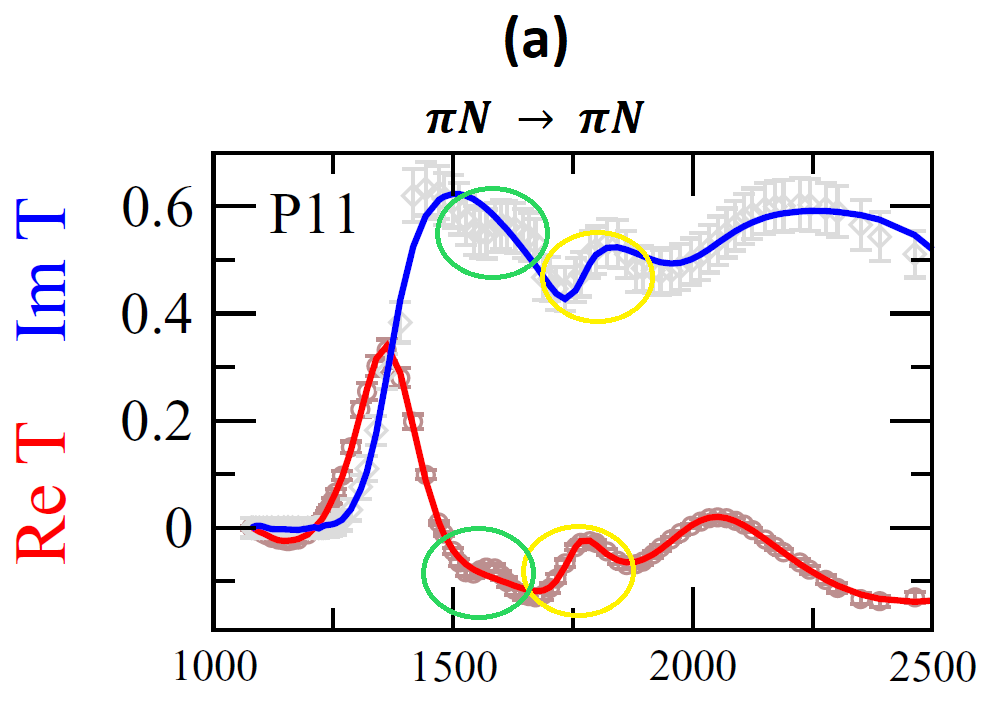}
 \includegraphics[width=0.45\textwidth]{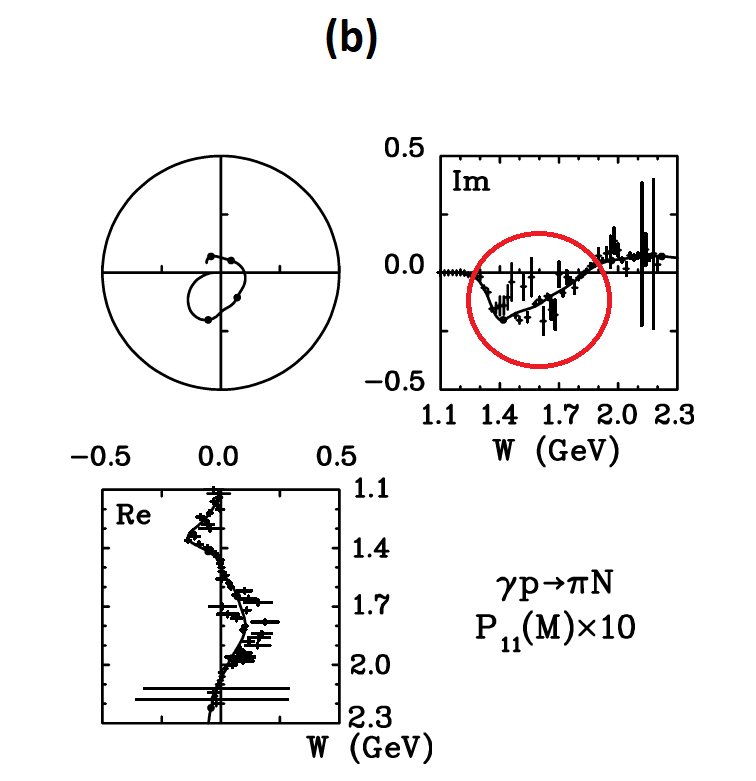} \\
  \includegraphics[width=0.45\textwidth]{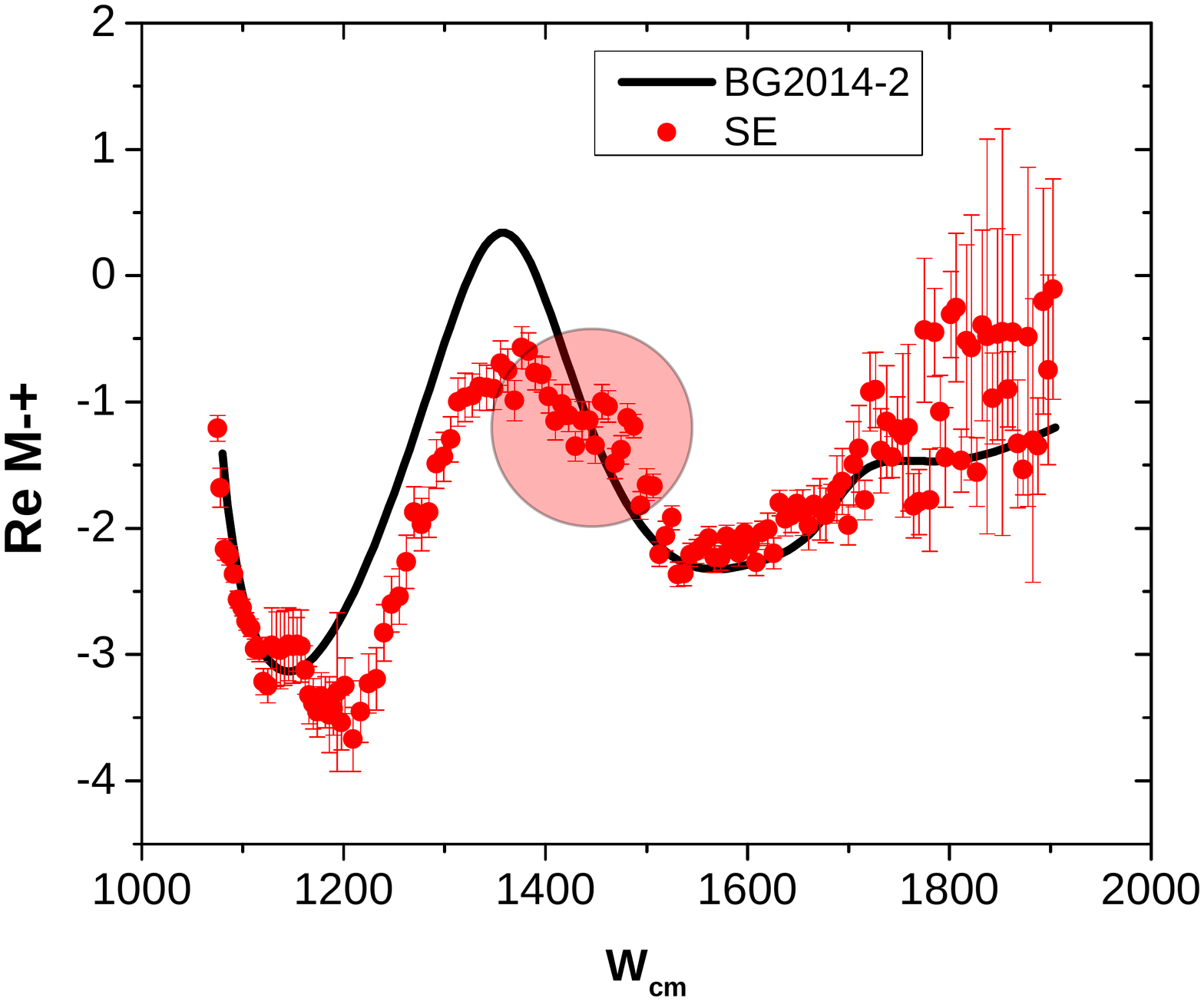}
   \includegraphics[width=0.45\textwidth]{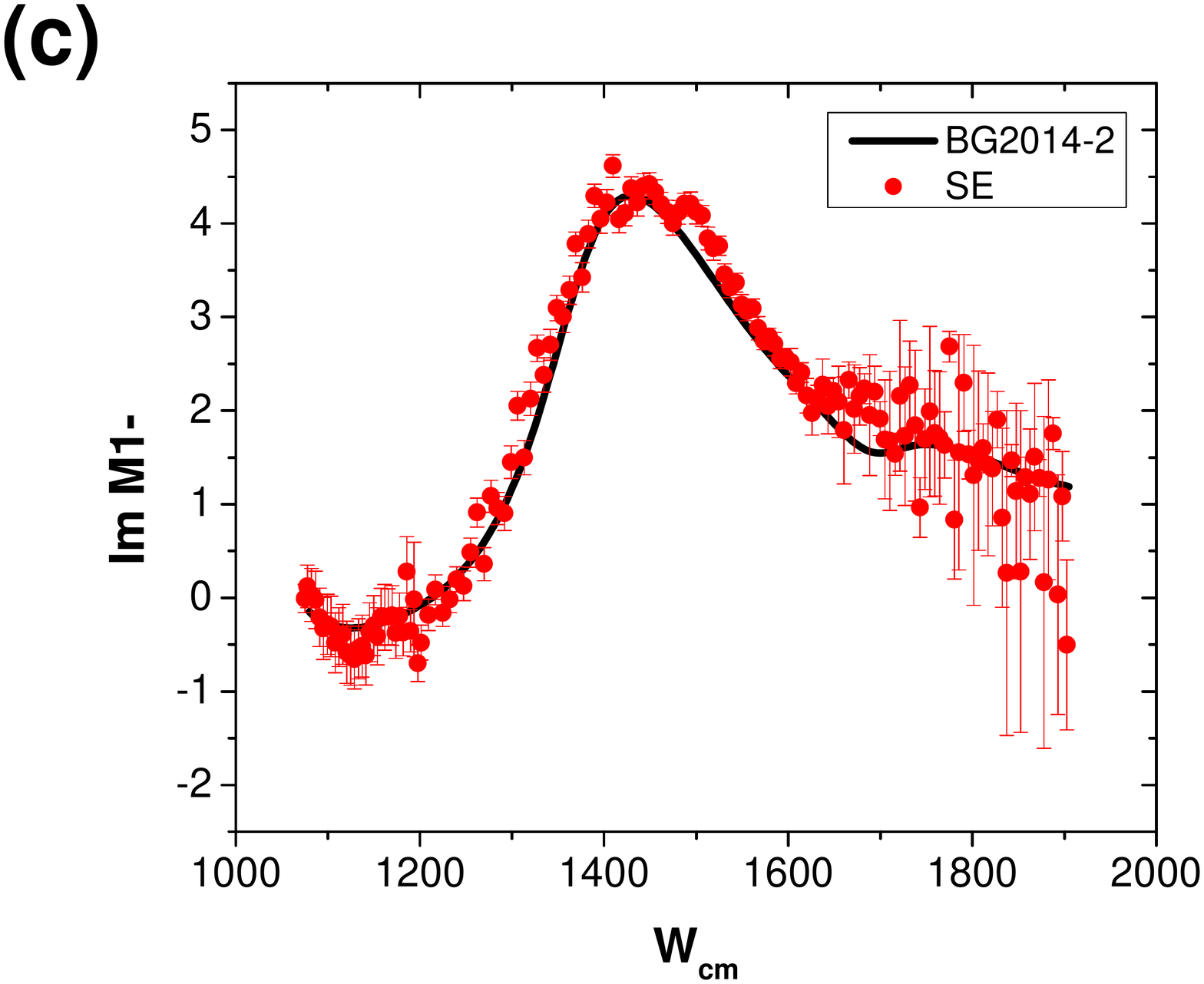}
\caption{\label{Fig7}(Color online) Signs for the possibility of a new pole in $P_{11}$ $1/2+$ partial wave in other processes.     }
\ec
\end{figure}

As can be seen from Table~\ref{Table1}, the fit with only the standard N(1710) 1/2+ state is poor  for this partial wave when compared to the rest. However, if we allow for one extra pole in the fit, the result is drastically improved, and the $\chi^2$ drops from 106 to 25. The difference in fits is shown in Fig.(\ref{Fig6}).
 The needed pole falls too far away to be interpreted as the known N(1440) 1/2+ state, but is pretty reliable. It has a well defined mass of 1526 $\pm$ 25 MeV and width of 73  $\pm$ 37 MeV with a reasonable residue.
  \newpage
  The obvious reason to doubt this state is that it has not been seen in other processes. However, one is allowed to speculate that this is not entirely correct. In Fig.(\ref{Fig7}) we give you the relevant partial wave for other processes: $\pi N$ elastic scattering from KH80 \cite{Hoehler84,Svarc2014}, SE part of $\pi N$-photoproduction from ref.~\cite{HuntManley} and $\pi^0$ photoproduction from ref.~\cite{Osmanovic2019}.  First it is important to stress that this state is never seen in ED models. This is not surprising, as to get a resonance in the ED model you have to put it in by hand in some way. As this resonance was never discussed, it was never added to the ED model, so it cannot be found there. This is also not surprising as strong, direct, experimental indications were never found.  However, some renown SE models show some problems (discontinuities, larger uncertainties) exactly on that energy. First, $\pi N$ elastic scattering KH80 \cite{Hoehler84,Svarc2014}. This is a fixed-t analysis, so by default it is an energy independent one. It is depicted in Fig.~(\ref{Fig7}-a). You see that it contains two distinct structures, one identified as N(1440) 1/2+ at low energies, and the second at high energies above 2000 MeV identified as N(2100) 1/2+. However, in addition  one sees two energy ranges of a notable departure from linearity: one at $\approx 1700$ MeV (indicated with yellow circle), but one also around 1550 MeV (indicated by the green circle).  Historically, there had for quite some time been a strong dispute about the effects at the energy range of 1700 MeV. GWU/SAID PWA claimed smooth behaviour without any poles, while KH80 \cite{Hoehler84} and some coupled-channel analyses \cite{Batinic,Ceci} identified it with N(1710) 1/2+ resonance. The dispute was solved in favor of N(1710) 1/2+ when inelastic-channel reactions  like $\pi N \rightarrow K \Lambda$  and $\pi N \rightarrow \eta N$ were included in coupled-channel formalisms~\cite{Batinic,Juelich}. However, the second area of non-linearity around 1550 MeV (marked with green circles) was never discussed. Another SE PWA for $\pi$-photoproduction~\cite{HuntManley} depicted in   Fig.~(\ref{Fig7}-b) also shows continuity in ED solution, but definite departure from linearity in this energy range (indicated with red circle) is observed. The third process is fixed-t SE $\pi^0$-photoproduction analysis depicted in Fig.~(\ref{Fig7}-c). Here the SE analysis  also shows notable departures from smooth, ED model, exactly in that range (also indicated with red circles). Neither effects have ever been explained.
\newpage
 Let me conclude:  analyzed few body processes of $\pi N$ elastic scattering, $\pi$ and $\eta$-photoproduction show the disturbance in that energy range, and our results in $\eta$-photoproduction strongly suggest a new resonant state in that range which couples dominantly to inelastic channels. So, it will be very interesting to see if such effect will be  confirmed or refuted  when  other inelastic photoproduction data like $K \Lambda$ data are also included in SE analyses.

\newpage
\section{Conclusions}
Analytic structure of Sol 1 of ref.~\cite{Svarc2020} is unclear in spite of the fact  that it fits the data almost perfectly. Following the theoretical arguments of ref.~\cite{Svarc2018} we attributed these effects to the small, but still uncontrolled phase change of reaction amplitudes. To prove this hypothesis we constrained the phase much stronger to the theoretical ED phase of ref.~\cite{BoGa}, and generated a new solution Sol 1/21. As expected, the agreement of the fit with the data was somewhat spoiled ($\chi^2$ was worse but still acceptable), but L+P analysis showed that the analytic structure (pole content) of the solution is much improved. Poles are clearly determined, and all of them are within one standard deviation consistent with the most recent compilation in PDG.  The only exception is $P_{11}$ 1/2+ partial wave where the need for additional analytic structure in the solution was strongly needed. In spite that it could be interpreted as some kind of uncontrolled threshold behaviour, we speculate that it might be the sign of a new, low lying $P_{11}$ 1/2+ state ($M = 1526 \pm 25$  MeV and $\Gamma = 73  \pm 37$ MeV). This state was never seen before in any SE PWA, but unclear and completely uninterpreted discontinuities in that energy range in $\pi N$ elastic, and $\pi$-photoproduction were reported \cite{Hoehler84,Svarc2014,HuntManley, Osmanovic2019}. The unification of all troubles with accepting a new state is offered as the simplest explanation\footnote{If it looks like a duck, swims like a duck, and quacks like a duck, then it probably is a duck}. Residues, as discussed in the text, are much less confident, and must be correlated with the phase used in the model.
\\ \\ \indent
This confirms that AA/PWA model of ref.~\cite{Svarc2020} in addition to reproducing the data almost perfectly has in addition a correct analytic structure, so it makes it a reasonable tool to be applied for a SE PWA in other processes. In the presently analyzed process of $\eta$-photoproduction strong S-wave dominance pushes the rest of the multipoles strongly down, so imprecision of currently available data base strongly influences their shape. We expect that much more stable situation in the case of $K \Lambda$-photoproduction what is the next target process will improve the situation and refine the insight into the preset problems.
\\ \\ \indent
It has been demonstrated that the good analytic structure of the model fails if the input reaction-amplitude phase is violated in order to improve  the agreement of the result with the data which is not ideal if the input phase is strongly enforced. The needed improvement in $\chi^2$ in AA/PWA  can be accomplished only by changing the phase; and this results in destruction of clear analytic structure. This indicates that the presently used BG2014-2 phase is good, but it is still not perfect. There is definitely quite some room for further improvements, but strictly within the framework of coupled-channel formalism to avoid continuum ambiguity effects.
\\ \\ \indent
Even small change of the angular part of the reaction amplitude phase can dramatically change the analytic structure (pole content) of obtained partial waves, so any change of the phase must be done in the controlled way, strongly correlated with other channels. As in any inelastic, single-channel model because of continuum ambiguity reaction-amplitude phase cannot be determined from the first principles, it has to be taken from somewhere, usually a coupled-channel model.  Free change of the phase in any SE PWA is not allowed.  So, each SE PWA must be model dependent from the first principles.
\acknowledgements{Many thanks go to Lothar Tiator from Institut f\"{u}r Kernphysik, Universit\"{a}t Mainz, my coworker and a good colleague who spent a lot of time in discussing different aspects of the paper, and double-checking some results. I indeed regret that he didn't join the project to the very end. I also acknowledge a great help from Yannick Wunderlich from Institut f\"{u}r Kernphysik, Universit\"{a}t Mainz who helped a lot in technical aspects of setting up long and complicated codes in  Mathematica. I would also like to acknowledge the long lasting support of  Johhanes Guttenberg University of Mainz, and in particular to Institut f\"{u}r Kernphysik what made the progress of this research much faster. This project has received funding from the European Union’s Horizon 2020 research and innovation programme  STRONG2020-HaSp under grant agreement No 824093.
   }


\newpage

\begin{thebibliography}{AA}
\bibitem{Svarc2018} A. \v{S}varc, Y. Wunderlich, H. Osmanovi\'{c}, M. Had\v{z}imehmedovi\'{c}, R. Omerovi\'{c}, J. Stahov, V. Kashevarov, K. Nikonov, M. Ostrick, L. Tiator, and R. Workman, Phys. Rev. \textbf{C 97}, 054611 (2018).
\bibitem{Svarc2020} A. \v{S}varc, Y. Wunderlich, and L. Tiator, Phys. Rev. \textbf{C 102}, 064609 (2020).
\bibitem{BoGa} A.~V.~Anisovich {\it et al.},  Phys.\ Rev.\ C\ {\bf 96}, 055202 (2017) and references therein,  https://pwa.hiskp.uni-bonn.de/.
\bibitem{Svarc2016} A. \v{S}varc, M. Had\v{z}imehmedovi\'{c}, H. Osmanovi\'{c}, J. Stahov,  L. Tiator, and R. Workman, Phys. Lett. \textbf{B 755}, 452-455. (2016), and references therein.
\bibitem{Hoehler84}
        G. H\"{o}hler,
        \emph{Pion Nucleon Scattering}, Part 2, Landolt-Bornstein:
        Elastic and Charge Exchange Scattering of Elementary Particles, Vol. 9b (Springer-Verlag, Berlin, 1983).
\bibitem{Osmanovic2018}
H. Osmanovi\'{c}, M. Had\v{z}imehmedovi\'{c}, R. Omerovi\'{c}, J. Stahov, V. Kashevarov, K. Nikonov, M. Ostrick,
L. Tiator, and A. \v{S}varc, Phys. Rev.\textbf{ C 97}, 015207 (2018).
\bibitem{Osmanovic2019}
H. Osmanovi\'{c}, M. Had\v{z}imehmedovi\'{c}, R. Omerovi\'{c}, J. Stahov,  M. Gorchtein,  V. Kashevarov, K. Nikonov, M. Ostrick,
L. Tiator, and A. \v{S}varc, Phys. Rev.\textbf{ C 100}, 055203 (2019).
\bibitem{Juelich} D. R\"{o}nchen, M. D\"{o}ring, H. Haberzettl, J. Haidenbauer, U. -G. Meissner and K. Nakayama Eur. Phys. J.\textbf{ A 51}, 70 (2015),
and references therein; and http://collaborations.fz-juelich.de/ikp/meson-baryon/main.
\bibitem{GWU/SAID} R. L. Workman, R. A. Arndt, W. J. Briscoe, M. W. Paris, and I. I. Strakovsky, Phys. Rev.\textbf{ C 86}, 035202 (2012); and
http://gwdac.phys.gwu.edu/.
\bibitem{MAID} D. Drechsel, S. S. Kamalov and L. Tiator, Eur. Phys. J.\textbf{ A 34 }(2007) 69; and https://maid.kph.uni-mainz.de/.
\bibitem{PDG}M. Tanabashi et al. (Particle Data Group), Phys. Rev.\textbf{ D 98}, 030001 (2018) and 2019 update.
\bibitem{Svarc2014}  A. \v{S}varc,  M. Had\v{z}imehmedovi\'{c}, H. Osmanovi\'{c}, R. Omerovi\'{c}, and J. Stahov, Phys. Rev.\textbf{ C 51}, 2310-2325 (1995).
\bibitem{HuntManley}B. C. Hunt and D. M. Manley, Phys. Rev.\textbf{ C 99}, 055205 (2019).
\bibitem{Batinic}M. Batini\'{c}, I. \v{S}laus, A. \v{S}varc, and B.M.K Nefkens, Phys. Rev.\textbf{ C 51}, 2310-2325 (1995).
\bibitem{Ceci} S. Ceci, A. \v{S}varc, and B. Zauner, Phys. Rev. Lett. \textbf{97}, 062002 (2006).

\end{thebibliography}
\end{document}